%% file: main.tex
\documentclass[]{aa}
\usepackage[utf8]{inputenc}
\usepackage{graphicx}
\usepackage{ulem,color}
\usepackage[dvipsnames]{xcolor}
\usepackage{array}   
\usepackage{multirow}
\usepackage{comment}

\newcommand{\qso}{HE\,0001$-$2340}
\newcommand{\zabs}{\ensuremath{z_{\rm abs}}}
\newcommand{\avg}[1]{\left< #1 \right>} 
\newcommand{\zem}{\ensuremath{z_{\rm em}}}

\newcommand{\lya}{\rm \mbox{Ly-$\alpha$}}
\newcommand{\Lya}{\rm \mbox{Ly-$\alpha$}}

\newcommand{\CII}{\ion{C}{ii}}
\newcommand{\CIV}{\ion{C}{iv}}
\newcommand{\CaI}{\ion{Ca}{i}}
\newcommand{\CaII}{\ion{Ca}{ii}}

\newcommand{\DI}{\ion{D}{i}}
\newcommand{\FeI}{\ion{Fe}{i}}
\newcommand{\FeII}{\ion{Fe}{ii}}

\newcommand{\HI}{\ion{H}{i}}
\newcommand{\MgI}{\ion{Mg}{i}}
\newcommand{\MgII}{\ion{Mg}{ii}}

\newcommand{\NaI}{\ion{Na}{i}}

\newcommand{\OI}{\ion{O}{i}}

\newcommand{\SII}{\ion{S}{ii}}
\newcommand{\SiII}{\ion{Si}{ii}}
\newcommand{\SiIV}{\ion{Si}{iv}}

\newcommand{\AlII}{\ion{Al}{ii}}
\newcommand{\AlIII}{\ion{Al}{iii}}
\newcommand{\kms}{\ensuremath{{\rm km\,s^{-1}}}}
\newcommand{\cmsq}{\ensuremath{{\rm cm}^{-2}}}

\newcolumntype{L}{>{$}l<{$}} 

\newcommand{\iap}{Institut d'Astrophysique de Paris, CNRS-SU, UMR\,7095, 98bis bd Arago, 75014 Paris, France \\ \email{noterdaeme@iap.fr}\label{iap}}
\newcommand{\eso}{European Southern Observatory, Alonso de C\'ordova 3107, Vitacura, Casilla 19001, Santiago 19, Chile \label{eso}}
\newcommand{\ioffe}{Ioffe Institute, {Politekhnicheskaya 26}, 194021 Saint Petersburg, Russia \label{ioffe}}
\newcommand{\uchile}{Departamento de Astronom\'ia, Universidad de Chile, Casilla 36-D, Santiago, Chile \label{uchile}}
\newcommand{\dgac}{Direction Générale de l'Aviation Civile, CRNA Nord, 9 rue de Champagne, 91200 Athis-Mons, France \label{dgac}}
\newcommand{\unige}{Department of Astronomy, University of Geneva, Chemin Pegasi 51, 1290 Versoix, Switzerland \label{unige}}

\title{Sharpening quasar absorption lines with ESPRESSO\thanks{Based on observations carried out at the Very Large Telescope of the European Southern Observatory under Prog. ID 0102.A-0463(A).}}
\subtitle{Temperature of warm gas at $z\sim2$, constraints on the Mg isotopic ratio, and structure of cold gas at $z\sim0.5$ }
\titlerunning{A VLT/ESPRESSO investigation of the line-of-sight to \qso}
\author{
P.~Noterdaeme\inst{\ref{iap}},
S.~Balashev\inst{\ref{ioffe}},
C.~Ledoux\inst{\ref{eso}},
G.~Duchoquet\inst{\ref{dgac}}, 
S.~L{\'o}pez\inst{\ref{uchile}},
K.~Telikova\inst{\ref{ioffe}},
P.~Boiss\'e\inst{\ref{iap}},
J.-K.~Krogager\inst{\ref{iap}},
A.~De~Cia\inst{\ref{unige}},
J.~Bergeron\inst{\ref{iap}}
}
\authorrunning{P. Noterdaeme et al.}
\institute{\iap \and \ioffe \and \eso \and \dgac \and \uchile \and \unige}

\date{\today 
}
\abstract{}
{We aim at studying several key physical properties of quasar absorption-lines systems that are
subtly encoded in their absorption profiles, but have been little investigated and/or are poorly constrained
so far.}
{We analyse a high-resolution ($R=140\,000$) spectrum of the bright quasar \qso\ ($\zem=2.26$), obtained with
ESPRESSO, recently installed at the Very Large Telescope. We analyse three systems at $z=0.45$, $z=1.65$, and $z=2.19$ using multiple-component Voigt-profile fitting. We also compare our 
spectrum with those obtained with VLT/UVES, covering a total period of 17 years.}
{We disentangle turbulent and thermal broadening in many components spread over about 400~\kms\ in the $z\approx 2.19$ sub-damped Lyman-$\alpha$ system. We derive an average temperature of 16000$\pm$1300~K, i.e., about twice the canonical value
of the warm neutral medium in the Galactic interstellar medium. A comparison with other high-$z$, low-metallicity absorbers reveals an anti-correlation between gas temperature and total \HI\ column density.
Although requiring confirmation, this could be the first observational evidence of a thermal
decrease with galacto-centric distance, i.e., we may be witnessing a thermal transition between the circum-galactic medium and the cooler ISM. 
We revisit the Mg isotopic ratios at $z=0.45$ and $z=1.65$ and constrain them to be
$\xi = (^{26}$Mg+$^{25}$Mg)/$^{24}$Mg~$<0.6$ and $<1.4$ in these two systems, respectively. These values are consistent
with the standard Solar ratio, i.e., we do not confirm strong enhancement of heavy isotopes previously inferred
from UVES data.
Finally, we confirm the partial coverage of the quasar emission-line region by a \FeI-bearing cloud in the $z=0.45$ system
and present evidence for velocity sub-structure of the gas that has Doppler parameters of the order of only $\sim 0.3$~\kms. This
agrees well with the low kinetic temperature of $T\sim 100$~K inferred from modelling of the gas physical conditions.
}
{This work demonstrates the uniqueness of high-fidelity, high-resolution optical spectrographs on large telescopes as tools
to investigate the thermal state of the gas in and around galaxies as well as its spatial and velocity structure on small scales,
and to constrain the associated stellar nucleosynthetic history.}


\begin{document}

\maketitle

\section{Introduction}
The advent of high-resolution spectrographs on 8-10~m class telescopes, in particular the High Resolution Echelle Spectrometer \citep[HIRES,][]{Vogt1994} on the Keck telescope followed by the Ultraviolet and Visual Echelle Spectrograph \citep[UVES,][]{Dekker2000} on the Very Large Telescope (VLT), has played a crucial role in the exploration of the distant Universe in absorption towards quasars and Gamma-ray burst afterglows. These spectrographs have enormously increased not only the number and variety of absorption-line systems observed at high spectral resolution but also the amount of information that can be extracted from them.
High-resolution spectroscopy has allowed fine understanding of the diffuse gas, in particular its chemical enrichment \citep[e.g.][among many others]{Prochaska2007, Petitjean2008,Vladilo2018, DeCia2018, Peroux2020}, its kinematics \citep{Prochaska1997,Ledoux2006}, as well as its physical state and the prevailing physical conditions \citep[e.g.][]{Srianand2005, Vreeswijk2007, Neeleman2015, Noterdaeme2017, Balashev2017}. 

Through the years, the astrophysical community has learned to decode wealth of  
information available in the absorption spectra, with other applications on a wide range of topics, from the possible space-time variation of fundamental constants (e.g. \citealt{Milakovic2021} and references therein), constraints on the sizes of the background source emission line regions \citep{Balashev2011, Bergeron2017}, measurements of the cosmic microwave background temperature at high-$z$ \citep{Noterdaeme2011}, measurements of the primordial abundance of deuterium \citep[e.g.][]{Cooke2014}, etc. Many of these studies, which rely on analysing fine details in the spectra, have reached a point where progress is mostly limited by the data. 
With a resolution up to several times that of other spectrographs, together with an exquisite wavelength calibration, the Echelle SPectrograph for Rocky Exoplanets and Stable Spectroscopic Observations \citep[ESPRESSO,][]{Pepe2021} on the Very Large Telescope (VLT) has the potential to reveal a new level of details. By getting closer to the intrinsic line widths, which are in general well below the resolution element of other spectrographs, ESPRESSO opens the possibility to investigate the structure of cold gas with very narrow lines, to break the degeneracy between macroscopic (turbulent) and microscopic (thermal) motions in gaseous clouds, and to investigate the contribution of different isotopes to a given absorption line, among other applications.

This paper deals with three different and independent topics that all take advantage of the same 
ESPRESSO spectrum: that of the quasar \qso. We therefore chose a less standard paper structure. After describing
the observations and data reduction in Sect.~\ref{s:obs} and the systems present 
along the line-of-sight in Sect.~\ref{s:sys}, the following sections deal with the temperature of the diffuse gas at $z=2.2$ (Sect.~\ref{s:DLAT}), the Mg isotopic ratio at $z=0.45$ and $z=1.65$ (Sect.~\ref{s:Mgiso}), and the structure of the cold gas at $z=0.45$ (Sect.~\ref{s:PC}), which we briefly introduce at the beginning of each section. We summarise our results in Sect.~\ref{s:concl}.

\section{Observations and data reduction \label{s:obs}}

\input{tab_obslog}

The observations of \qso\ ($V=16.7$, $z_{\rm em}=2.28$) were carried out in service mode
during the nights of November 9, 12, and 28, 2018, using ESPRESSO in single UT mode at
the Melipal Unit Telescope \#3. Located at the incoherent combined Coud\'e focus of the VLT,
ESPRESSO is a highly-stabilised, thermally-controlled \'echelle spectrograph fed by two
fibres, one for the target and the other for simultaneous reference calibration. We recorded
the sky light in Fibre B for a proper sky subtraction. We used the High-Resolution (HR) mode of the
instrument leading to a median resolving power of 139,000 as measured from emission lines
in ThAr calibration frames. The fibre aperture on sky in this mode is $1\farcs 0$. Ambient
conditions while observing were good with a typical seeing of $0\farcs 89$, at an average
airmass of 1.13, and clear sky. Once inside the spectrograph, the light from the two fibres
is dispersed by an \'echelle grating and the spectral orders are split up into a Blue
(379-525~nm) and a Red (525-788~nm) channel. The corresponding spectra are recorded on
separate Charge Coupling Devices whose pixels were binned two by one in the spatial
direction. This provides a spatial sampling per slice of 4.5~pixels, and similar sampling
in the dispersion direction.

The data were reduced using the ESO Reflex workflow of the ESPRESSO Data Reduction System (DRS)
version 2.0.0 publicly released by ESO. After bias, dark, and inter-order
background subtraction, an optimal extraction of the 2D spectral orders is performed using
master flat-fields as order profiles. During this process, dead, hot, and saturated pixels
are masked out and cosmic-ray hits are rejected. After flat-fielding and de-blazing, the 2D
spectra were wavelength-calibrated using the combination of ThAr and Fabry-P\'erot light
sources. The sky spectrum was extracted from Fibre B and subtracted from the science frames. The
latter were then rebinned and merged into 1D data products with associated error and quality
maps. Flux calibration was performed using estimated absolute efficiencies and
atmospheric extinction curves. 
{Each exposure was corrected for atmospheric absorption features by fitting a synthetic transmission spectrum to the Red-arm spectra using the public code {\tt molecfit} \citep{Smette2015, Kausch2015}.}
Individual exposures were then combined with a drizzling-like
approach using the espda\_coadd\_spec recipe of the Data Analysis System (DAS) v1.0.3
after conversion of the wavelength scales to helio-vacuum. The resolving power adopted for
each of the final Blue and Red-arm spectra is given in Table~\ref{tab:obslog}. It is
about 5\% higher (resp. lower) in the Red (resp. the Blue) compared to the overall
median value. {The average signal-to-noise ratio per pixel is about $S/N\sim 25$ in the Blue and 
$S/N\sim 45$ in the Red.}

As we will complement ESPRESSO with UVES data to investigate the possible time-variation of
absorption lines in Sect.~\ref{s:PC}, we provide a log of all these observations in
Table~\ref{tab:obslog}. Details on the UVES data are presented by \citet{Dodorico2007},
\citet{Agafonova2011} and \citet{Bergeron2017}. {The S/N per pixel of the UVES spectra 
are around 90, 65 and 60 at 420~nm for the combined 2001, 2009 and 2017 spectra, 
respectively\footnote{{Note that the four spectra (ESPRESSO, UVES-2001, UVES-2009 and UVES-2017) 
will be jointly analysed in Sect.~\ref{s:PC}, but are not combined altogether.}}}.

\section{Systems along the line-of-sight to \qso \label{s:sys}}

Capitalising on the high quality of our ESPRESSO spectrum, we performed a
systematic identification of absorption-line systems covered by the observations. 
{Firstly, we visually identified the most obvious absorption systems based on the easily recognisable doublets \MgII$\lambda\lambda$2796,2803 and \CIV$\lambda\lambda1548,1550$ and marked absorption features from other species at the same redshift. We then scanned visually the full spectrum, identifying the remaining absorption features until no unidentified feature remained. This procedure was done independently by two team members (PN and GD) and re-checked by a third member (CL).} We made our best effort to identify even weak features
that may be helpful for other studies than presented here.
We used
both the combined 1D spectrum and the individual exposures to ascertain the astrophysical
nature of features versus possible telluric residuals, which are shifted differently when
applying the heliocentric correction. 

In total, we identify no less than twelve absorption systems with redshifts ranging between
$z=0$ and the quasar emission redshift, $z_{\rm em}=2.26$. These are summarised in
Table~\ref{t:systems} together with the detected transition lines.

\input{systems}

\section{Kinetic temperatures of the gas at $z=2.2$ \label{s:DLAT}}

Thermal balance in neutral gas leads to an equation of state with two stable phases
under an external pressure \citep{Field1969}: a warm and diffuse phase (a.k.a. the warm neutral medium; WNM)
with temperatures of the order of 10$^{4}$~K and a cold, denser phase (the cold neutral
medium; CNM) with $T\sim 100$~K. Below some minimal pressure, the gas is mostly a WNM
while it is predominantly a CNM above some maximal pressure. Between these
two pressure limits, as is typically the case in galactic discs, the gas is a
mixture of both warm and cold phases \citep[e.g.,][]{Wolfire2003}. Other, more unstable phases
also complete this picture \citep[e.g.,][]{Salpeter1976}, but for simplicity,
in the following we stick to a canonical two-phase description.

The temperature and phase mixing of the neutral gas have been investigated locally using
H\,{\sc i} 21-cm emission/absorption data \citep[e.g.,][]{Heiles2003} but this technique
is still inapplicable to the distant Universe. At high redshift, the neutral gas is currently 
only detectable through damped Lyman-$\alpha$ absorption imprinted on the spectra of
bright background sources.

The small incidence rate of 21-cm absorption \citep[e.g.,][]{Kanekar2014} and molecular
hydrogen \citep[e.g.,][]{Ledoux2003,Noterdaeme2008b,Jorgenson2014,Balashev2018} --both
sensitive probes of cold gas-- as well as more indirect constraints from
the fine-structure excitation of \CII\ and \SiII\ \citep{Neeleman2015} show that most DLAs
actually probe the WNM (see also \citealt{Krogager2020b} for estimations of intrinsic
statistics, i.e., corrected for selection effects). While the temperature of the CNM is
well constrained from the excitation of the low rotational levels of H$_2$
\citep[see, e.g.,][among other studies]{Ledoux2002,Noterdaeme2007,Balashev2019},
the temperature of the WNM at high redshift is generally \textit{assumed} to be around
the canonical, local value (8000\,K). This has been used to derive the CNM fraction from
the observed harmonic mean of the 21-cm spin temperature \citep[e.g.,][]{Srianand2012}.
In fact, direct measurements of the WNM temperature are difficult to obtain and essentially
rely on determining the thermal broadening of the lines that results from
the Maxwell--Boltzmann distribution of gas particle velocities. However, since
macroscopic motions are also present, the actually observed Doppler broadening ($b$) is
the quadratic sum of thermal and turbulent broadening:
$b^2 = b_{\rm th}^2 + b_{\rm turb}^2 = {2 k_{\rm B} T}/{m} + b_{\rm turb}^2$, where $T$ is the
gas temperature, $m$ the mass of the species, and $k_{\rm B}$ the Boltzmann constant.
Disentangling thermal from turbulent broadening then requires observations of several
species with different masses within the same clouds, but this also depends on
precise measurements of total $b$ which are hard to obtain even at relatively high S/N and
spectral resolution \citep{Carswell2012}.

\subsection{Thermal broadening}

The availability of ESPRESSO is an excellent opportunity to disentangle turbulent and
thermal broadening \citep{Lee2020} over the full extent of QSO absorption-line profiles,
in particular when light species are well constrained. In the case of the DLA
at $z=2.187$ towards \qso, the \CII\ and \OI\ absorption-line complexes exhibit unsaturated
components that make the measurement easier to perform. \citet{Richter2005} already performed
a detailed analysis of this system using UVES data. We therefore do not wish to redo a complete study of
this system but rather focus on disentangling thermal from turbulent broadening. For
our analysis, we used the VPFIT software package v12.3 \citep{Carswell2014}.

As a starting point, we used the velocity structure inferred by \citeauthor{Richter2005} as a
first guess in the fitting process and we have adopted their component labelling for an easier comparison.
We do not see evidence for component $F$, which is rejected by VPFIT, while at the
higher resolution of ESPRESSO the modelling of component A requires splitting into
three sub-components, that we here denote $A_{1}$, $A_{2}$, and $A_{3}$ (see Table~\ref{t:DLAmet}). 

\begin{figure*}
    \centering
    \includegraphics[clip=,trim=0.0cm 4.3cm 2cm 0.0cm, width=0.9\hsize]{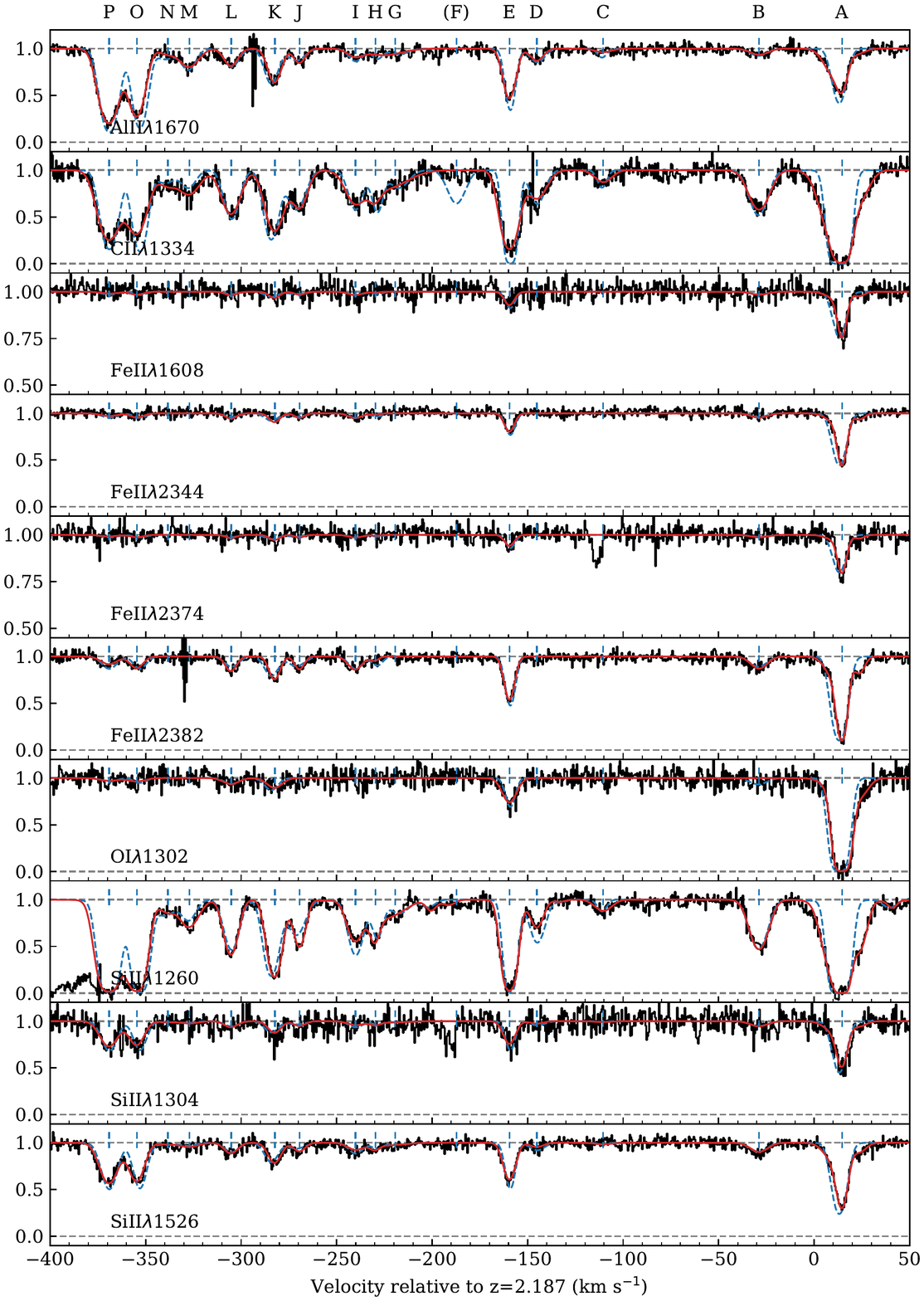}
    \caption{Multi-component Voigt-profile fitting of metal lines in the sub-DLA
    at $z=2.187$. The normalised ESPRESSO spectrum is shown in black, with
    the best-fit synthetic spectrum overlaid in red. The blue dashed curve corresponds to
    the synthetic spectrum computed using the best-fit parameters obtained
    by \citet{Richter2005} from the analysis of UVES data. Note that their component $A$ here
    splits into three sub-components that we call $A1$, $A2$, and $A3$
    in Table~\ref{t:DLAmet} and their component $F$ at $v\sim -150$~\kms\ appears to be
    spurious as seen from \CII$\lambda$1334.
    \label{f:DLAmet}}
\end{figure*}

We performed the fit allowing for a mixture of turbulent and thermal broadening in
each component.
The best-fit results are shown in Fig.~\ref{f:DLAmet} and the corresponding parameter values
are listed in Table~\ref{t:DLAmet}. For most components (namely 12 out of 17), the degeneracy
between turbulent and thermal broadening is easily broken and we obtained good constraints on
$b_{\rm turb}$ and $T$. We note that, for component $D$, thermal broadening dominates over
turbulence and the fit is essentially consistent with $b_{\rm turb} =0~\kms$.
However, several components (i.e., $C$, $G$, $H$, $M$, and $N$) have poorer constraints and
their fit is consistent with turbulent broadening alone (i.e., the $1\sigma$ temperature
range encompasses $T=0$~K) and therefore in this case we instead report upper limits on the temperature. The weakest
components ($G$ and $N$) are very close to the noise level so that even the total
Doppler parameter values are almost unconstrained, which leads to loose constraints on
the temperature that are of little practical use, if any.

\input DLAmet_newfwhm

To further test the robustness of the decomposition into thermal and turbulent contributions, we
also fitted the data using independent (total) $b$-parameters for each species and derived
$b_{\rm turb}$ and $T$ \textit{a posteriori} from 
{the relation between the total Doppler parameter squared and 
} the inverse of the mass of each species as done by, e.g., \citet{Carswell2012}. 

\begin{figure}
    \centering
    \includegraphics[clip=,trim=0cm 1.5cm 0cm 2.0cm,width=0.85\hsize]{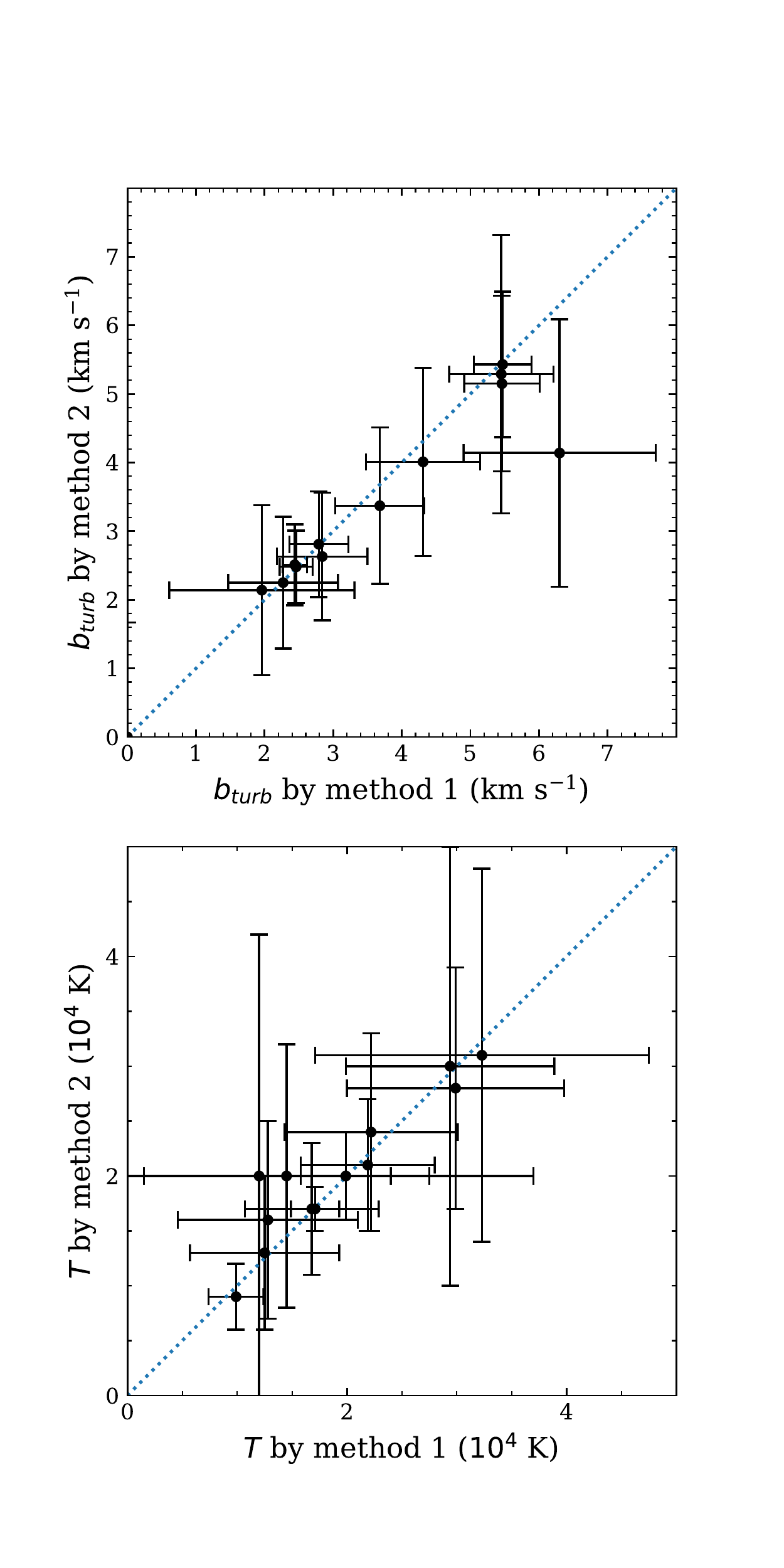}
    \caption{Comparison of the turbulent component of the Doppler parameter (top) and the
    temperature (bottom) derived from two different methods (1: using $b_{\rm turb}$ and 
    $T$ as free parameters for all species of a given component; 2: fitting total $b$-values
    for each species and deriving $b_{\rm turb}$ and $T$ a posteriori from the variation
    of $b$ with the inverse of the species mass). Blue dotted lines show the
    one-to-one relation. 
    }
    \label{f:bTcomp}
\end{figure}

{Although the errors on the Doppler parameters evaluated independently for each species in each component become large, the increase of $b$-values with decreasing mass of the species, following the relation $b \propto m^{-0.5}$, is very well determined. Such a measurement was possible for fourteen components, which we show in Fig.~\ref{f:T_by_meth_2}.}
One advantage of this method is that any
departure from the linear relation for a given species would help identify an issue with the
data or the fitting process, or could in principle mean that the species do not pertain all to the same gas.
A comparison between turbulent Doppler parameters and temperatures obtained with each of the two methods
is shown in Fig.~\ref{f:bTcomp}. Unsurprisingly, they are very much consistent with
each other, hence supporting the convenient (and standard) assumption of co-spatiality,
allowing us to use turbulent plus thermal broadening for each velocity component while fitting (i.e. method 1). {This also permits the use of fewer free parameters while the total reduced $\chi_{\nu}^2$ remains unchanged. In the following, we therefore use the values derived from method 1, i.e. with physically motivated link between Doppler parameters of different species, but the results remain unchanged if values from method 2 were used instead.}

\begin{figure}
    \centering
    \includegraphics[clip=,trim=0.8cm 0cm 1.5cm 0.5cm,width=\hsize]{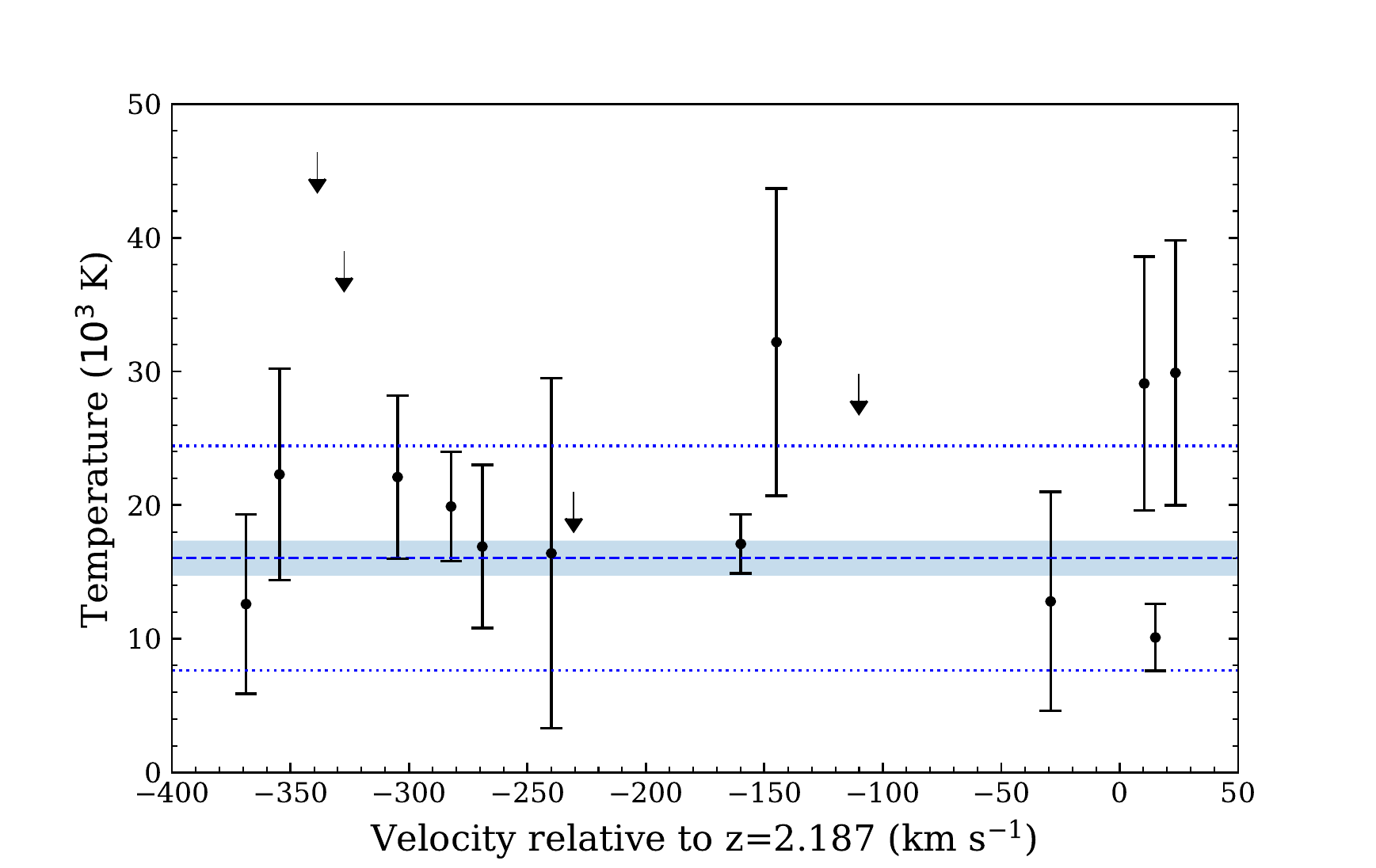}
    \caption{Kinetic temperatures determined in various components along the line profile
    of the sub-DLA at $\zabs=2.187$ (zero of the velocity scale). The dashed blue line and
    shaded area correspond to the weighted mean and associated error
    $\avg{T}=16000 \pm 1300$~K. The dotted lines show one standard deviation around
    this mean. 
}
    \label{f:Tbcomp}
\end{figure}

We measure temperatures in the range $9000-32000$~K (see Fig.~\ref{f:Tbcomp}) with a
dispersion of about 6300~K, which is of the same order as the $1\sigma$ uncertainty
on individual temperature measurements. These values are indeed consistent within uncertainties with a
single temperature for all of the velocity components of the system, with weighted mean
$\avg{T}=16000 \pm 1300$~K (unweighted $\avg{T}=20000$~K, excluding upper limits only). This temperature is twice as high as the canonical value
of 8000~K generally assumed for the Galactic ISM \citep{Wolfire2003}.

\subsection{Comparison with other measurements at $z>2$}

To investigate whether the higher-than-canonical temperature measured here is inherent to
one system or is more generally seen in high-$z$ DLAs, a sample of DLA temperature
measurements must be compiled.
In fact, to our knowledge, no systematic study of this type has ever been performed. One possibility is to use deuterium, which is the second lightest species after hydrogen and whose $\sim 10^{5}$ times smaller abundance enables the observation of unsaturated \DI\ lines.

Little attention has yet been paid to this possibility \citep[but see][]{Noterdaeme2012a}
mainly because the focus was on measuring the primordial D/H ratio
\citep[e.g.,][]{Cooke2018} where the measurement of the temperature is only a by-product. We
therefore searched the literature for additional temperature measurements, in particular
in \DI-related publications which are easy to identify. When the temperatures were
not explicitly given {(for the systems towards Q0913+072 and J1558-0031)}, we derived them from the provided 
$b$-values {\citep{Pettini2008,OMeara2006}}. Because of the
proximity between \DI\ and \HI\ lines ($\Delta v = -81.6$~\kms), and because
the systems where generally selected to be metal-poor (so that the gas is less
processed and the D/H ratio is close to primordial), these measurements concern only one or
a few components at the blue edge of the profiles. We also included the measurement
of \citet{Dutta2014} who emphasised the importance of taking into account thermal broadening
to discuss the abundance of Carbon in metal-poor DLAs.\footnote{\citet{Milakovic2021}
recently used the HARPS spectrograph and derived gas temperatures as by-products in a study
of the possible cosmic-time variation of the fine-structure constant. Since the
redshift ($\zabs=1.1$) and metallicity ($\log Z/Z_{\rm \odot}=-0.22$) of this system
are different from the rest of our sample, we do not include it in our analysis. We note however
that the corresponding mean temperature ($9100\pm1100$~K) would perfectly match the
anti-correlation discussed here.}

\begin{figure*}
    \centering
    \includegraphics[clip=,trim=0.5cm 0.0cm 1cm 0.5cm,width=\hsize]{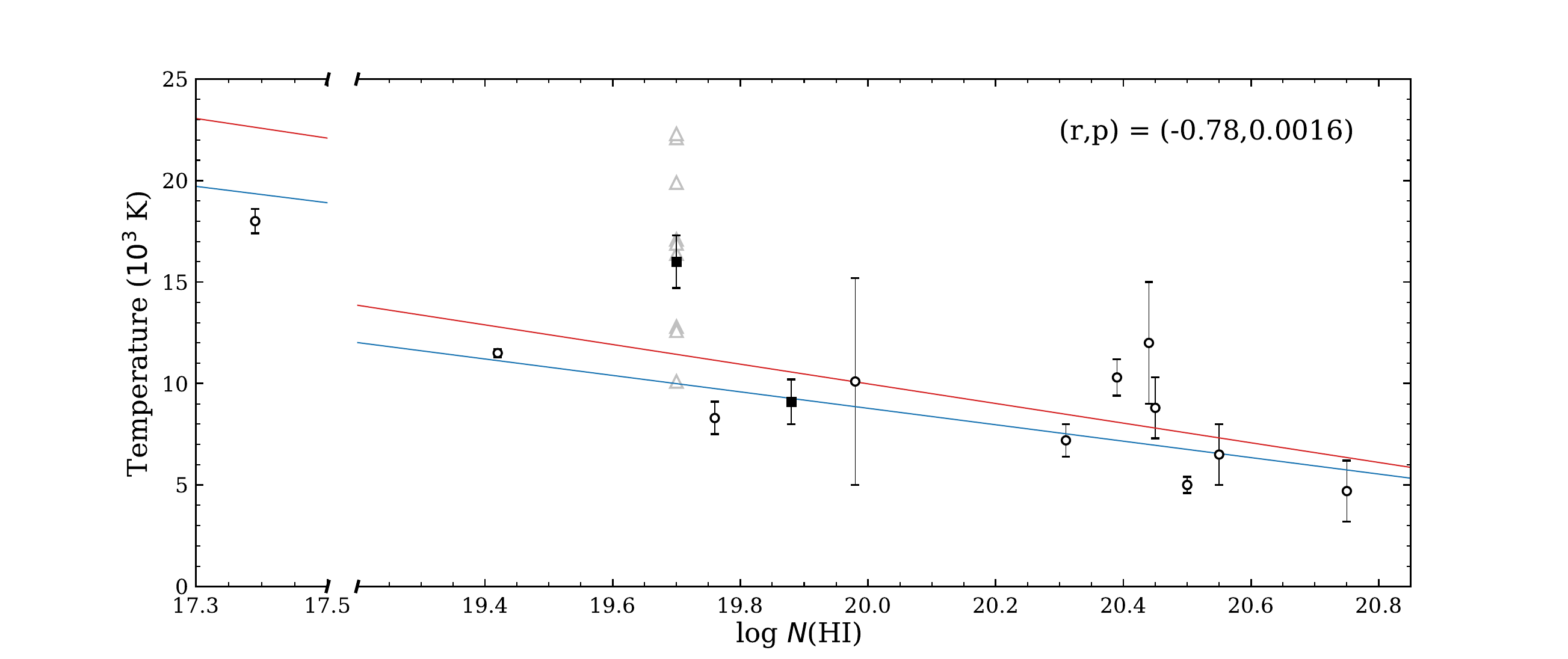}
    \caption{Kinetic temperatures derived from Doppler broadening in various absorption-line
    systems as a function of the {\sl total} \HI\ column densities of these systems.
    Empty circles correspond to measurements obtained thanks to the presence of deuterium
    (typically in one or a few close components) while filled squares correspond to average
    measurements in complex systems made possible by high-resolution spectroscopy.
    Grey triangles represent measurements in individual components of the DLA
    towards \qso\ (Fig.~\ref{f:Tbcomp}).
    The data exhibit an anti-correlation with Pearson coefficient $r=-0.78$
    and probability that this is pure chance coincidence $p=0.16\%$. 
    The solid blue (respectively red) line corresponds to an unweighted linear fit including
    (resp. excluding) the lowest $N(\HI)$ value towards Q\,1009$+$2956.
    }
    \label{f:density_T_corr}
\end{figure*}

The compiled temperature measurements are summarised in Table~\ref{f:Temps}.
All systems are DLAs or sub-DLAs, with the exception of the absorber at $z=2.504$
towards Q\,1009$+$2956 \citep{Burles1998} which is a LLS in which most of the gas is
likely ionised. Fig.~\ref{f:density_T_corr} reveals an anti-correlation between
the kinetic temperature and the total \HI\ column density of the systems, with
Pearson correlation coefficient $r = -0.78$ and a probability of a chance coincidence of
$p = 1.6\times10^{-3}$.
A linear decrease of about 3500~K per decade of $N(\HI)$ does provide a reasonable
description of the data over the full range of $N(\HI)$. To test whether the
anti-correlation is driven by the only LLS, that also has the highest temperature in our
sample, we repeated the exercise without that data point and obtained $r = -0.63$, with
a probability that this is due to chance coincidence of still only about 3\%. 
In this case, the decrease in temperature with increasing $N(\HI)$ is even steeper with 5000~K
per $N(\HI)$-dex.

\subsection{On the anti-correlation between $T$ and overall $N(\HI)$}

The decrease in kinetic temperature with increasing $N(\HI)$ appears to be real
but the physical reason behind it is not straightforward to understand. Importantly, temperatures are
measured \textit{locally}, generally in a few components, while the measured \HI\ column
density corresponds to the total value throughout the absorber. This indicates that while the
kinetic temperatures are likely determined by conditions belonging to individual clouds
(e.g., the strength of the UV field, ionisation fraction, and gas metallicity), it also shows some
dependence on larger scales.
The metallicities are low in our sample, and the systems may therefore correspond to
gas located far away from star-forming regions so that the UV flux is dominated by
the meta-galactic background \citep[e.g.,][]{Khaire2019} and hence is similar for all
systems.
In that case, the observed $T$-$N(\HI)$ anti-correlation could be connected to the
known anti-correlation between \HI\ column density and impact
parameter \citep[e.g.,][]{Krogager2017}. One possibility is that, the lower the \HI\
column density the larger the distance from the galaxy, and the higher the mixing with
shock-heated ionised gas in the halo. Alternatively, the typically low $N(\HI)$ values in our
sample may enhance photo-ionisation heating and more generally alter thermal balance,
which is sensitive to the medium's ionisation fraction.
Unfortunately, we cannot discriminate between these explanations.
Further measurements of kinetic temperatures over a wide range of environments (ideally
with simultaneous constraints on impact parameters) and physical conditions are needed to
understand the observed $T$-$N(\HI)$ anti-correlation.

The dependence between a local (temperature) and a global property (total $N(\HI)$)
demands further investigation with quantitative modelling of the thermal state
of the gas. In addition to confirming the observed $T-N(\HI)$ anti-correlation, it will
be interesting to investigate its dispersion at any given column density. One could naively
expect that the dispersion is larger at higher column density --not only from one system to
another but also amongst components in a given system-- owing to varying
conditions when gas closer to or within galaxies is probed.
In turn, at large galacto-centric distances, conditions may vary more smoothly. Another
possibility is to investigate how the temperatures are related to metallicity and kinematics
as simulations suggest that outflowing gas has both higher metallicities and temperatures.
Deriving kinetic temperatures in a large sample of systems through high-resolution
spectroscopy may hence provide new, valuable clues to understanding the thermal state of the
neutral phase in the circum-galactic medium at high redshift.

 \begin{table*}
 \centering
 \caption{Kinetic temperatures in various systems determined from thermal broadening of the lines. 
 \label{f:Temps}} 
 \begin{tabular}{llllll}
 \hline\hline
 {\Large \strut}Quasar & \zabs  & $\log N(\HI)$     & $\log(Z/Z_{\rm \odot})$   &  $T$ ($10^{3}$\,K)          & Reference\\
 \hline
 Q1009+2956       & 2.504 &  17.39 $\pm$ 0.06   & ?                 & 18.0 $\pm$ 6.0  & \citet{Burles1998}      \\
 HS 0105+1619     & 2.536 &  19.42 $\pm$ 0.01   & -1.77             & 11.5 $\pm$ 0.2  & \citet{O'Meara2001}     \\
 \qso\            & 2.187 &  19.7               & -1.81 $\pm$ 0.07  & 16.1 $\pm$ 1.3  & This work               \\
 Q1243+307        & 2.525 &  19.76 $\pm$ 0.03   & -2.77 $\pm$ 0.03  & 8.3  $\pm$ 0.8  & \citet{Cooke2018}       \\
 J1444+2919       & 2.437 &  19.98 $\pm$ 0.01   & -2.04 $\pm$ 0.01  & 10.1 $\pm$ 5.1  & \citet{Balashev2016}    \\
 Q0913+072        & 2.618 &  20.31 $\pm$ 0.04   & -2.42 $\pm$ 0.01  & 7.2  $\pm$ 0.8  & \citet{Pettini2008}     \\
 J1419+0829       & 3.050 &  20.39 $\pm$ 0.01   & -1.92 $\pm$ 0.01  & 10.3 $\pm$ 0.9  & \citet{Pettini2012}     \\
 Q2206-199        & 2.076 &  20.44 $\pm$ 0.05   & -2.31 $\pm$ 0.07  & 12.0 $\pm$ 3.0  & \citet{Carswell2012}    \\
 CTQ 247          & 2.621 &  20.45 $\pm$ 0.10   & -1.99 $\pm$ 0.10  & 8.8  $\pm$ 1.5  & \citet{Noterdaeme2012a} \\
 J1358+6522       & 3.067 &  20.50 $\pm$ 0.01   & -2.33 $\pm$ 0.02  & 5.1  $\pm$ 0.4  & \citet{Cooke2014}       \\
 J0035-0918       & 2.39  &  20.55 $\pm$ 0.10   & -2.69 $\pm$ 0.17  & 6.5  $\pm$ 1.5  & \citet{Dutta2014}       \\
 J1558-0031       & 2.702 &  20.75 $\pm$ 0.03   & -1.65 $\pm$ 0.04  & 4.7  $\pm$ 1.5  & \citet{OMeara2006}      \\
 
 \hline
 \end{tabular}
 \tablefoot{
 The average temperatures and quoted uncertainties were either taken directly from the references provided in the last column, or computed from tabulated 
 values using the published temperatures or Doppler parameters. In some cases, no uncertainty were provided and the uncertainty provided here correspond to 
 the standard deviation between components. Because of this heterogeneity, we do not consider the uncertainties in our analysis (see text).
 }
 \end{table*}

\section{Constraints on the isotopic Mg ratio \label{s:Mgiso}}

Constraining the relative abundances of magnesium isotopes is a powerful probe of star formation processes over cosmological time scales since the main contributors to $^{24}$Mg are massive stars while the other two stable isotopes ($^{25}$Mg and $^{26}$Mg) come mostly from intermediate mass stars \citep{Vangioni2019}.
\citet{Agafonova2011} used a UVES spectrum of \qso\ to obtain the first measurement of the Mg isotopic ratio at cosmological distances, and derived a strong overabundance of heavy isotopes. 
The higher resolution achieved with ESPRESSO, better sampling, as well as the exquisite wavelength calibration allows us to revisit this measurement, in spite of the lower S/N \textit{per pixel} in our spectrum. In the following, we test two methods to measure this ratio. 

The first method corresponds 
to that used by \citet{Agafonova2011} and is based on the apparent velocity shift of \MgII\ lines. Indeed, for a given transition, the absorption line wavelengths from different isotopes are very similar, resulting in the observation of a single line. However, the apparent central rest-frame wavelengths of the lines does depend on the isotopic ratio so that comparing their apparent velocity shift with respect to other species can in principle constrain this ratio at any redshift \citep{Levshakov2009}. 

The second method relies on a direct fit, including explicitly the transitions from the three isotopes into the Voigt-profile model, assuming no intrinsic shift with respect to other species. 
This direct fit is possible since the isotopic wavelength separation is comparable to the ESPRESSO resolution element.

\subsection{Isotopic ratio from apparent velocity shifts}

\begin{figure}
    \centering
    \includegraphics[clip,trim=0.5cm 1cm 1.5cm 0.5cm, width=\hsize]{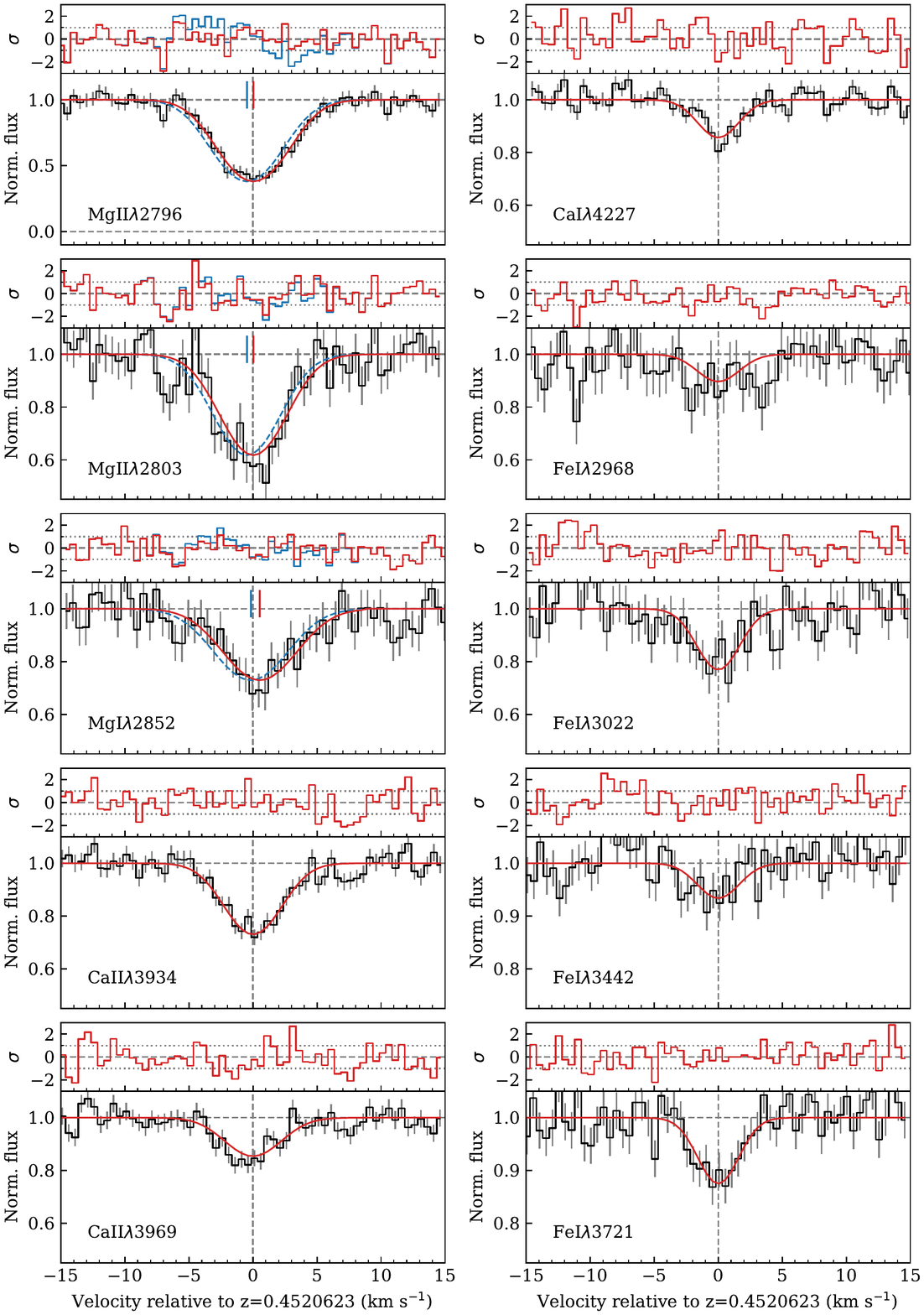}     
    \caption{Single Voigt profile fit to the cold, narrow component at $z=0.45206$ (ESPRESSO data: black with grey error bars, synthetic spectrum: red). Residuals (in units of standard deviation) are shown in the sub-panel above each line panel.  The zero of the velocity scale is taken at the velocity of the iron and calcium lines. The redshift of the magnesium lines were left vary independently from the other lines and are displayed as short red tick marks. The blue profiles and the corresponding blue tick marks and residuals correspond to the apparent velocity shift ($-0.47$ and $-0.17$ \kms\ from \MgII\ and \MgI\, respectively) measured by \citet{Agafonova2011} from UVES data.}
    \label{f:isotopic}
\end{figure}

In Fig.~\ref{f:isotopic}, we show the Voigt-profile fit to the \CaII, \CaI, \FeI, \MgII, and 
\MgI\ lines detected in the narrow component at $z=0.45206$. The redshift of
calcium and iron lines were tied together and taken as reference to compare the apparent velocities of the \MgII$\lambda\lambda$2796,2803 doublet 
and the \MgI$\lambda$2852 lines. The magnesium lines were fitted using a single component and the composite wavelengths from \citet{Murphy2014}, $\lambda_0=2796.35379,2803.530982$ (\MgII) and 2852.962797 (\MgI), that correspond to weighted wavelengths with Solar relative isotopic abundances 78.99:10:11.01 (${\rm ^{24}Mg:^{25}Mg:^{26}Mg}$).
From this, we obtained an apparent velocity shift of $+0.03 \pm 0.12$~\kms and $+0.53 \pm 0.32$~\kms, for \MgII\ and \MgI\, respectively. While the 
positive shift of \MgI\ indicates in principle an abundance of heavy isotopes less than Solar, this is only marginally significant (1.6\,$\sigma$) and based on a single and weak absorption line. From Fig.~\ref{f:isotopic}, it also appears that assuming a smaller velocity shift ($-0.17$~\kms\ from \citealt{Agafonova2011}) also provides a good fit, without noticeable structure in the residuals. Hence, we do not consider this line in the following anymore. The sensitivity of the \MgI\ line to the isotopic ratio is even smaller than for \MgII\ anyway.

In turn, the velocity shift for \MgII\ is better constrained but our measured value, $\Delta v (\MgII) = +0.03 \pm 0.12$~\kms, is in tension ($>3\,\sigma$) with the velocity shift, $\Delta v (\MgII) = -0.44 \pm 0.05$~\kms, obtained by \citet{Agafonova2011}\footnote{The authors use slightly different composite \MgII\ wavelengths, so that their velocity shift should in principle be increased by $\sim 0.03$~\kms\ when compared to our adopted values. The value for the adopted \MgI\ composite wavelength differs by only $\sim$ 2 m\,s$^{-1}$.}. This is clearly seen in Fig.~\ref{f:isotopic}, where assuming this latter value introduces a shift with respect to the ESPRESSO data, which is mostly noticeable as an $s$-shaped structure in the \MgII$\lambda$2796 residuals.

\begin{figure}
    \centering
    \includegraphics[clip=,trim=0.2cm 0cm 1cm 0.5cm,width=\hsize]{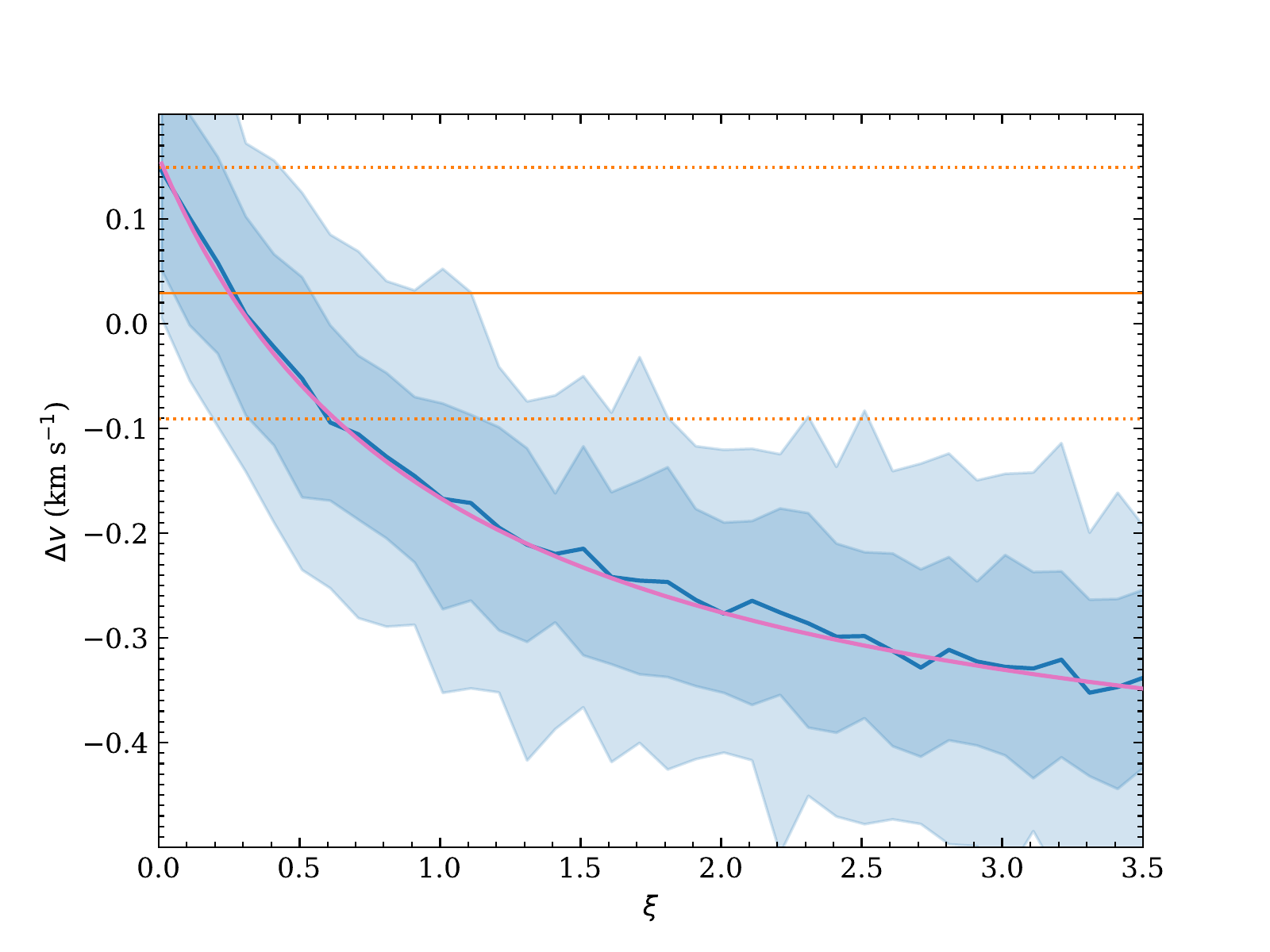}
    \caption{Velocity shift (with respect to lines with the standard isotopic ratio) as a function of $\xi = (^{25}$Mg$+^{26}$Mg$)/^{24}$Mg assuming $r = ^{26}$Mg$/^{25}$Mg~=~1 for mock ESPRESSO data. The solid blue line corresponds to the mean 
    value for 100 mock spectra for each input $\xi$ value, and the dark and light blue regions represent the regions containing 68.3\% and 95.5\% of the values, respectively. 
    The pink line corresponds to a noiseless mock spectrum.
    The observed velocity shift and associated 1$\sigma$ uncertainty for the system at $z=0.45206$ is shown as orange horizontal solid and dotted lines, respectively. 
    }
    \label{f:dv_xi}
\end{figure}

To test the significance of our measurement and the conversion from velocity shift to isotopic ratio, we generated artificial spectra featuring the \MgII\ absorption doublet with an input total column density of $N(\MgII)=10^{12}$~\cmsq\ to mimic the line observed here and varying isotopic ratio $\xi={\rm (^{25}Mg+^{26}Mg)/^{24}Mg}$, assuming 
$r={\rm ^{26}Mg/^{25}Mg} = r_{\rm \odot} \approx 1$. We convolved these spectra with the appropriate instrumental profile, used the same wavelength binning and added Gaussian noise (with S/N=20, as in our data). For each $\xi$ value, we generated 100 artificial spectra that we fitted using the same procedure. The results of this exercise are shown in Fig.~\ref{f:dv_xi}. 
The average obtained for 100 artificial spectra (dark blue line in Fig.~\ref{f:dv_xi}) agrees very well with the values obtained for a noiseless spectrum (pink line in the same figure), showing that the finite S/N of the data does not introduce any systematic bias.
Considering the \MgII\ values only, the measured velocity shift of $+0.03\pm0.12$~\kms\ translates to $\xi= 0.27^{+0.33}_{-0.27}$ (using the mean of 100 spectra with S/N=20).
The 68.3\% interval on the measured velocities in the artificial noisy data assuming this central $\xi$-value is $\approx 0.1$ km\,s$^{-1}$. This is in agreement with the formal 1\,$\sigma$ uncertainty from fitting the real data, and even slightly smaller. Indeed, the latter takes into account the uncertainty on the anchor redshift from the other lines in the error budget.

\subsection{Explicitly including the isotopes in the absorption model}

\begin{figure}
    \centering
    \includegraphics[clip=,trim=0.2cm 0cm 1cm 0.5cm,width=\hsize]{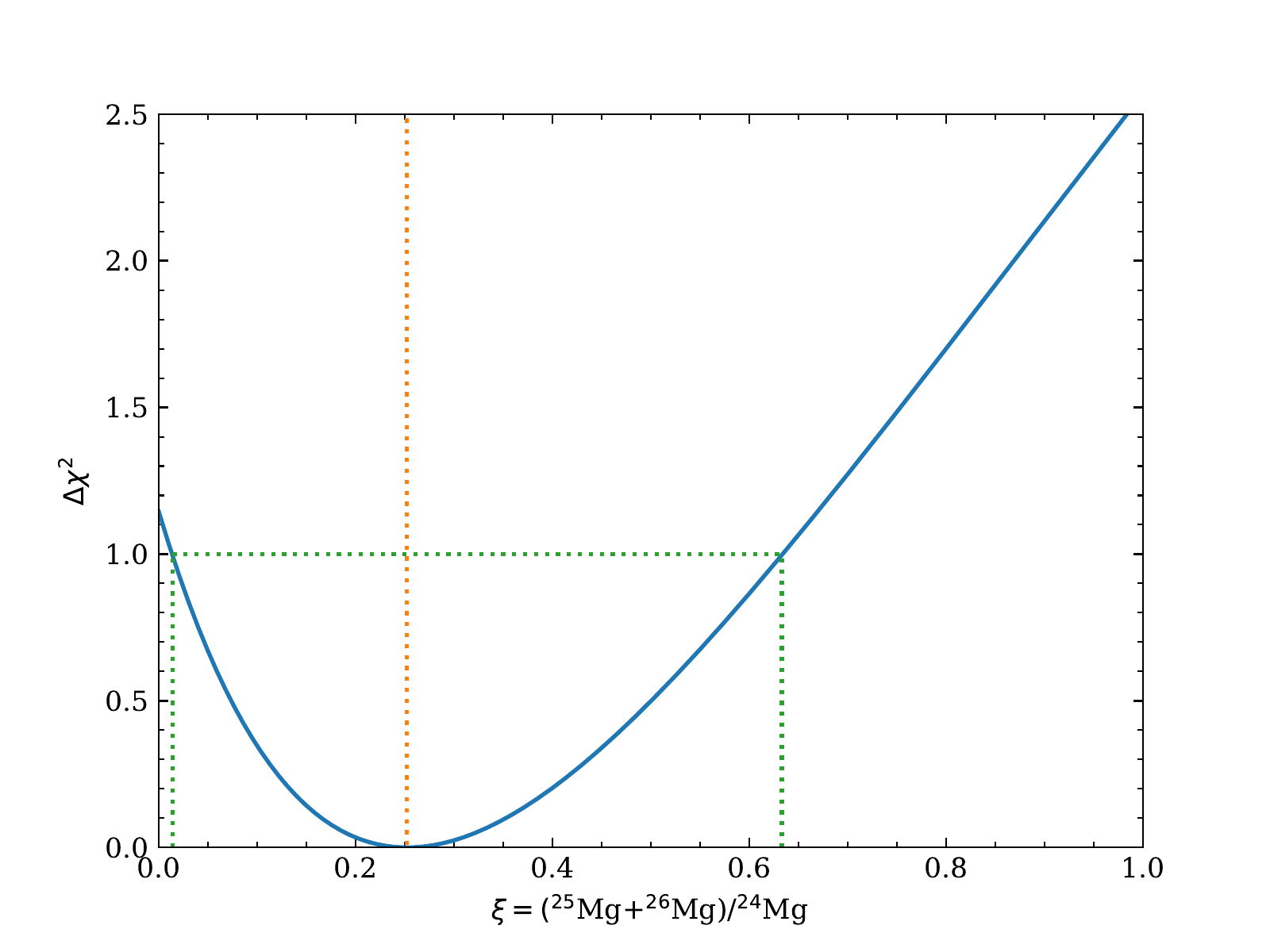}
    \caption{$\Delta \chi^2=\chi^2-\chi^2_{min}$ values as a function of the assumed isotopic ratio $\xi$ for the Voigt profile fit of \MgII\ lines including explicitly all three isotopes in the model. $\chi^2_{min}=899.6$ (891 degrees of freedom) is reached for $\xi=0.24$ (dotted orange line). The corresponding 1$\sigma$ range goes from $\xi=0.01$ to $\xi=0.63$ (green dotted lines). 
    } 
    \label{f:chixi}
\end{figure}

\begin{figure}
    \centering
    \includegraphics[clip=,trim=0.2cm 9.cm 11cm 0cm,width=0.8\hsize]{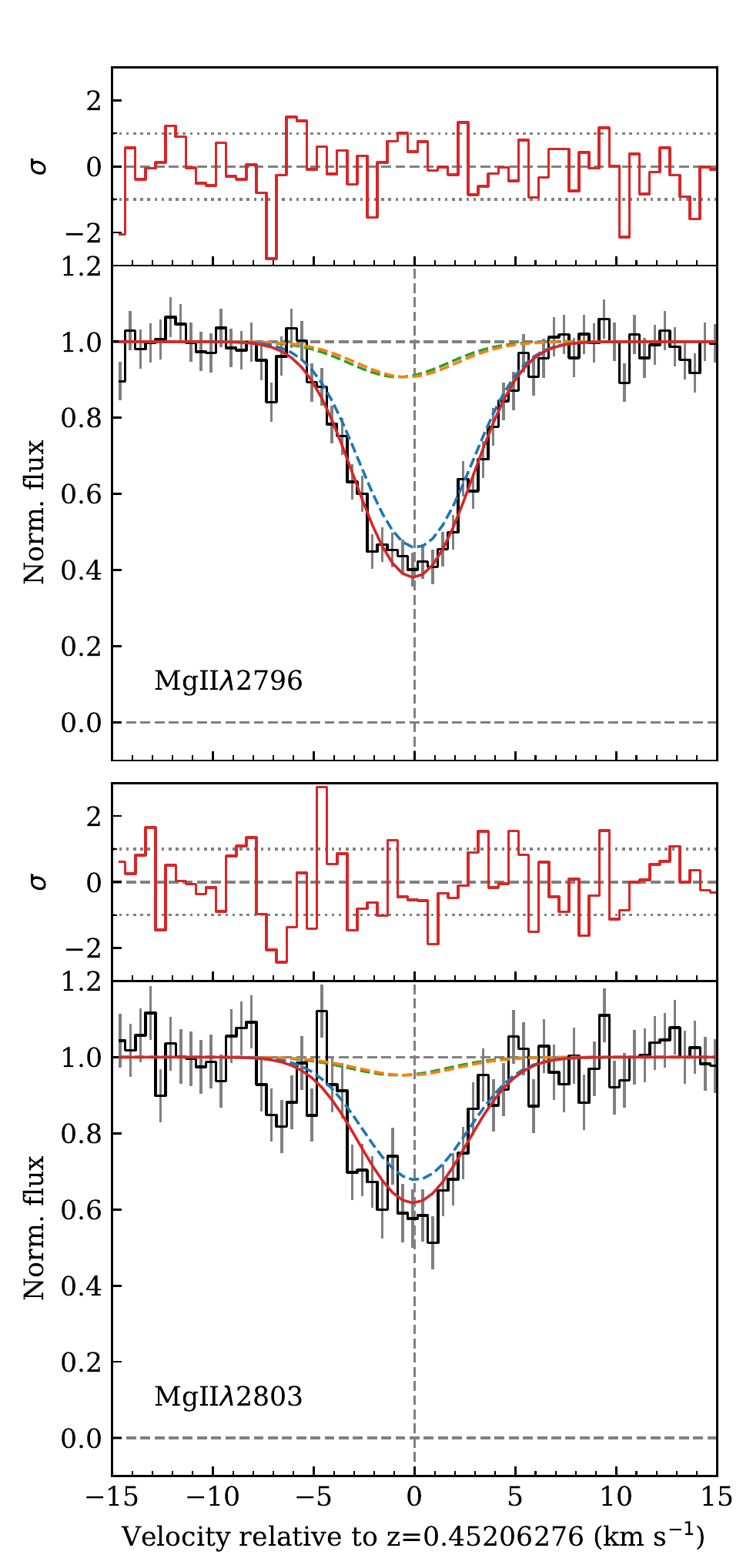} 
    \caption{Synthetic \MgII\ absorption profiles {for the system at $z=0.452$} (red line) corresponding to the best fit ($\xi=0.25$, assuming $r=1$) including $^{24}$\MgII, $^{25}$\MgII\ and $^{26}$\MgII\ lines (dashed-blue, dashed-orange and dashed-green, respectively) overlaid on the ESPRESSO data (black with grey error bars). The top panels show the residuals.} 
    \label{f:bestxi}
\end{figure}

Next, we tried the second, more direct method, including explicitly the three isotopes in the Voigt profile model and assuming no intrinsic shift with respect to other species.
We derived the best-fitting $\chi^2$ values for a range of assumed isotopic 
abundance ratio, following a standard method as described by \citet{Lampton1976}.
In practice, we tied together the redshift of all species shown in Fig.~\ref{f:isotopic} and assumed fixed isotopic ratio, while letting the total column densities and the Doppler parameters free. However, we assumed $r=^{26}$Mg/$^{25}$Mg = 1 as previously. The $\Delta\chi^2$ as a function of $\xi$ are shown in Fig.~\ref{f:chixi}. The minimum value is reached for $\xi=0.25$ (this best fit is shown in Fig.~\ref{f:bestxi}) and the associated 68.3\% confidence range using this single parameter is given by $\Delta\chi^2 = 1$. 
We then obtain $\xi=0.25^{+0.38}_{-0.24}$, nicely confirming the value and confidence region obtained from the apparent velocity shift method.

\subsection{The system at $z=1.65$} %
\begin{figure}
    \centering
    \includegraphics[clip=,trim=0.2cm 8.5cm 10cm 0.5cm, width=0.8\hsize]{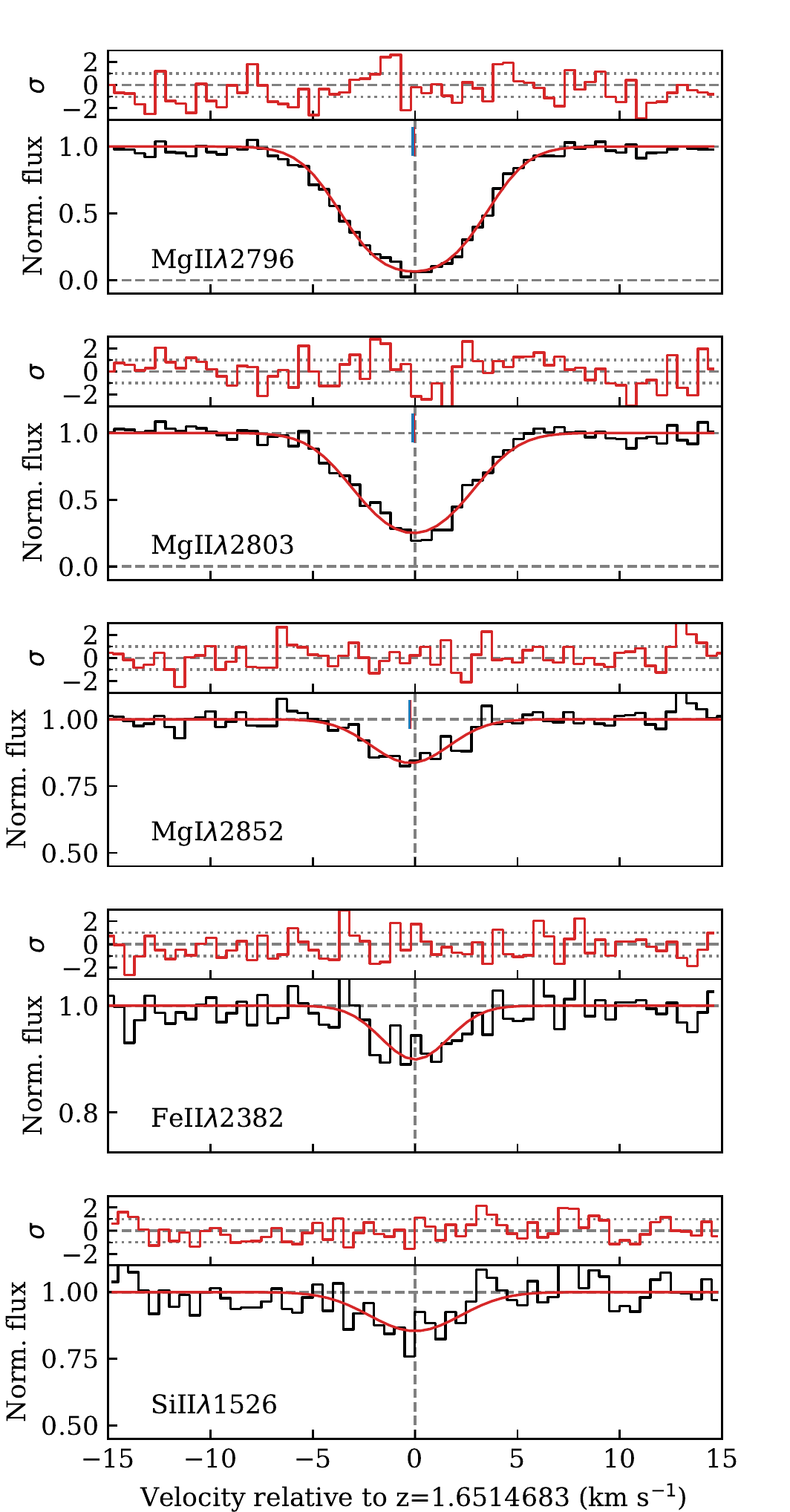}
    \caption{Fit to the metal lines at $z=1.65$. The long dashed line (zero of the velocity scale) correspond to the redshift of \FeII\ and \SiII, while the short ticks mark the centroids of the \MgII\ and \MgI\ lines. The velocity shifts are here indistinguishable from those measured from UVES data by \citet{Agafonova2011} for this system.}
    \label{f:isotopic1.6}
\end{figure}

\begin{figure}
    \centering
    \includegraphics[clip=,trim=0.2cm 0cm 1cm 0.5cm,width=\hsize]{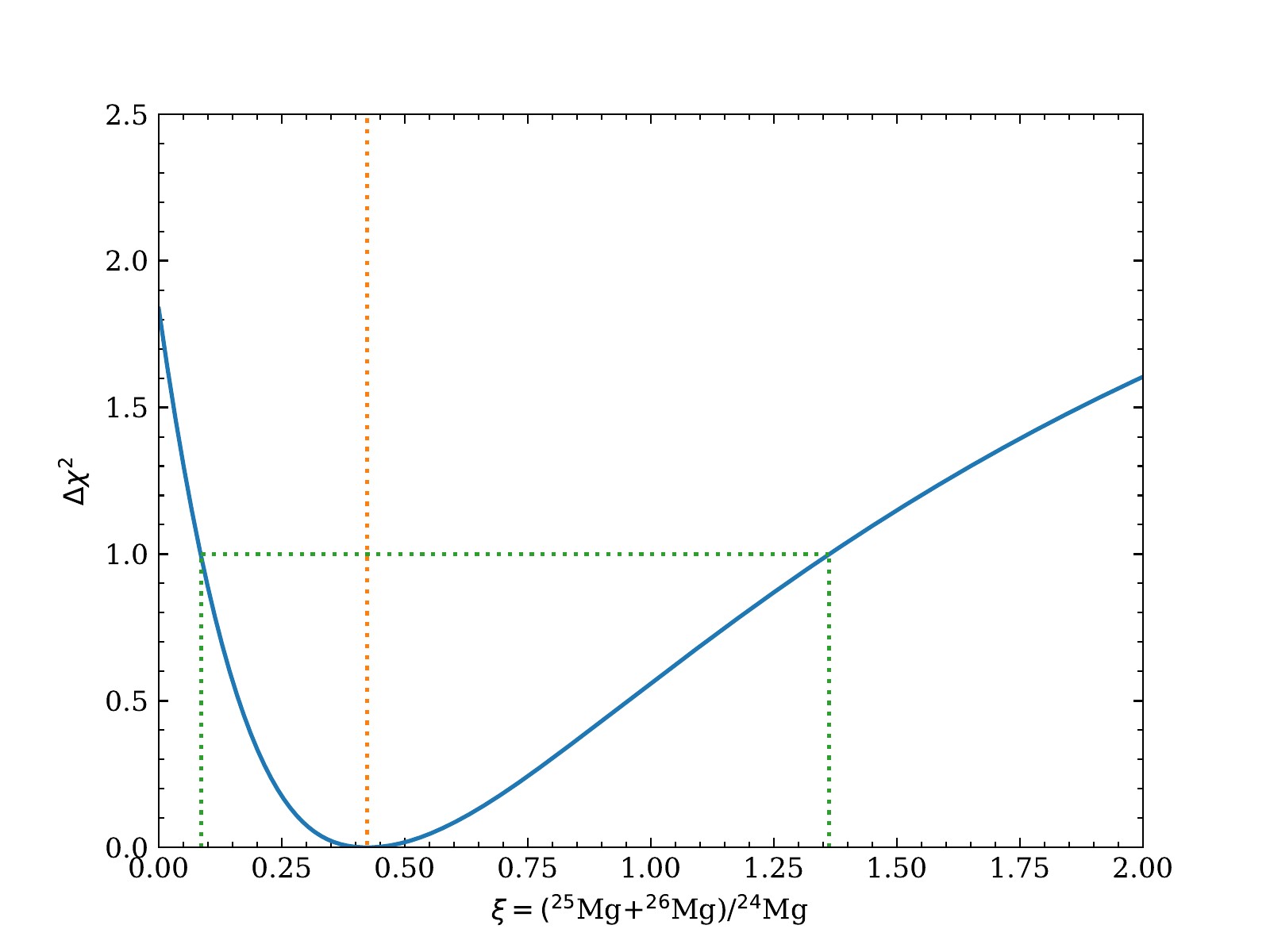}
    \caption{Same as Fig.~\ref{f:chixi} for the system at $z=1.65$.} 
    \label{f:chixic}
\end{figure}

\citet{Agafonova2011} have also obtained constraints for components in the systems at $z=1.58$ and $z=1.65$ from the same data along the same line of sight. For the system at $z=1.58$, modelling the absorption profile of the sub-system that they considered requires several components. In addition, the \MgII\ lines are saturated and partially affected by telluric absorption. Therefore, we did not attempt to get a constraint on the Mg isotopic ratio for this system. For the simpler, single component system at $z=1.65$, we obtain $\Delta v=-0.08 \pm 0.27$~\kms (\MgII) and  $\Delta v=-0.23 \pm 0.35$~\kms (\MgI), surprisingly consistent with the values reported by \citeauthor{Agafonova2011} ($-0.08 \pm 0.20$ and $-0.28\pm0.44$, respectively) and again consistent with no velocity shift {(see Fig.~\ref{f:isotopic1.6})}.  
Using the variation of $\chi^2$ as function of the assumed $\xi$ as described above, we derive a 68.3\% confidence interval of $\xi =0.09-1.36$ with the best fit-value at $\xi=0.42${, see Fig.~\ref{f:chixic}}.

\subsection{No evidence for strong enhancement of heavy isotopes}

Before concluding, we note that while our single component Voigt-profile fits provide a very good representation of the data for both systems, they may not capture any possible hidden velocity structure. Indeed, as we show in the next Section, sub-structure is likely present in the cold, \FeI-bearing gas.
Even in the presence of sub-structures, a strong enhancement of heavy isotopes remains unlikely.
Indeed, the presence of a velocity component in the red wing of the main profile could 
in principle compensate for the velocity shift induced by heavy isotopes, but it would require an ad-hoc combination of velocity and strength.
Furthermore, such a peculiar situation should be the case for both systems studied here. 
On the other hand, the presence of an additional component in the blue wing
of the profile could mimic the presence of $^{25}$Mg and $^{26}$Mg when their actual abundance would be close to zero. While we do not see any evidence for such a component in the profile of other species, the obtained constraints should conservatively be considered upper limits, i.e. $\xi<0.6$ and $\xi<1.4$ for the systems at $z=0.45$ and 1.65, respectively. We note that the obtained uncertainties on the $\xi$ values already allowed for zero abundance of heavy isotopes anyway.

Another issue concerns our assumption of $r={\rm ^{26}Mg/^{25}Mg} = 1 \approx r_{\odot}$. This is more difficult to address since allowing for a free value of $r$ would introduce more degeneracy which is hard to investigate with the S/N reached in our current data. In addition, these two heavy isotopes are expected to arise from stars of similar mass, so their relative abundance is not expected to vary much. With the exquisite wavelength calibration of ESPRESSO, it is not hopeless to investigate this issue, provided much higher S/N is obtained.

In conclusion, we do not confirm the strong enhancement of heavy Mg isotopes in any of the two systems considered here. The apparent shift observed in the UVES data is therefore likely neither due to a space-time variation of the fine-structure constant $\alpha$ \citep{Murphy2008} nor to strong enhancement of heavy Mg isotopes, but is instead due to subtle wavelength distortions in the UVES data, 
that have been revealed and investigated in a number of works aiming at precision spectroscopy \citep{Rahmani2013, Whitmore2015, Dumont2017}.

\section{Structure of the cold-gas absorber at $z=0.452$ \label{s:PC}}

{\citet{Dodorico2007} first identified the $z=0.45206$ absorption system and detected the neutral species \ion{Fe}{i}, \ion{Si}{i} and \ion{Ca}{i}, that are rarely found in quasar absorption systems.}
By modelling the physical conditions in {this}  
system, \citet{Jones2010} constrained the temperature to be $T<100$\,K and the density to be in the range $100-1000$\,cm$^{-3}$. 
Given the relatively small column densities inferred from their model ($N(\HI)$ in the range $10^{18.5-20.8}$~\cmsq), this suggests a small longitudinal size. For $N(\HI) = 10^{19.6}~\cmsq$ and a density of 300\,cm$^{-3}$, the size is $l_{\parallel} \sim $ 0.04 pc.

It is therefore reasonable to 
expect a similarly small transverse extent as well. 
A small transverse size of the cloud can in principle lead to several interesting effects. 
First, if at a given wavelength, the size of the background source becomes comparable to that of the absorbing cloud, then the former may not be considered a point source anymore. This leads to partial coverage \citep[e.g.][]{Balashev2011}. Indeed, \citet{Bergeron2017} found a partial coverage effect for the \ion{Fe}{I}$\lambda2719$\AA\ line that is by chance located on top of the Ly$\alpha$ emission line from the quasar. However, this \FeI\ line is blended with Ly$\alpha$ absorption, which complicates the analysis.  
Second, the presence of structures within the cloud may in principle lead to time variations of the line profiles since the exact location of the line of sight with respect to the absorbing gas may vary due to relative motions of the absorber and/or observer with respect to the background source \citep{Boisse2015}. However, the line of sight drifts are expected to be at best of the order of a few 10$^{-3}$~pc over 10 years.
This means that time variation of the profiles is expected only if significant structures are present at such extremely small scales. 

In order to constrain the transverse spatial and line-of-sight velocity structure of the absorber, we used our ESPRESSO spectrum together with UVES spectra obtained previously in 2001, 2009 and 2017, hence covering a time interval of 17 years. Unfortunately, the strongest \ion{Fe}{I} transitions at $2484$ and $2523$\AA\ are not covered by the ESPRESSO spectrum. Notwithstanding, they are covered by all UVES spectra that have bluer coverage. Since many \ion{Fe}{i} lines are weak, we took into account the uncertainty in continuum construction when fitting these lines. The fitting method is described in Appendix~\ref{s:feifit}.

\begin{figure}
    \centering
    \includegraphics[clip=,width=1.0\hsize]{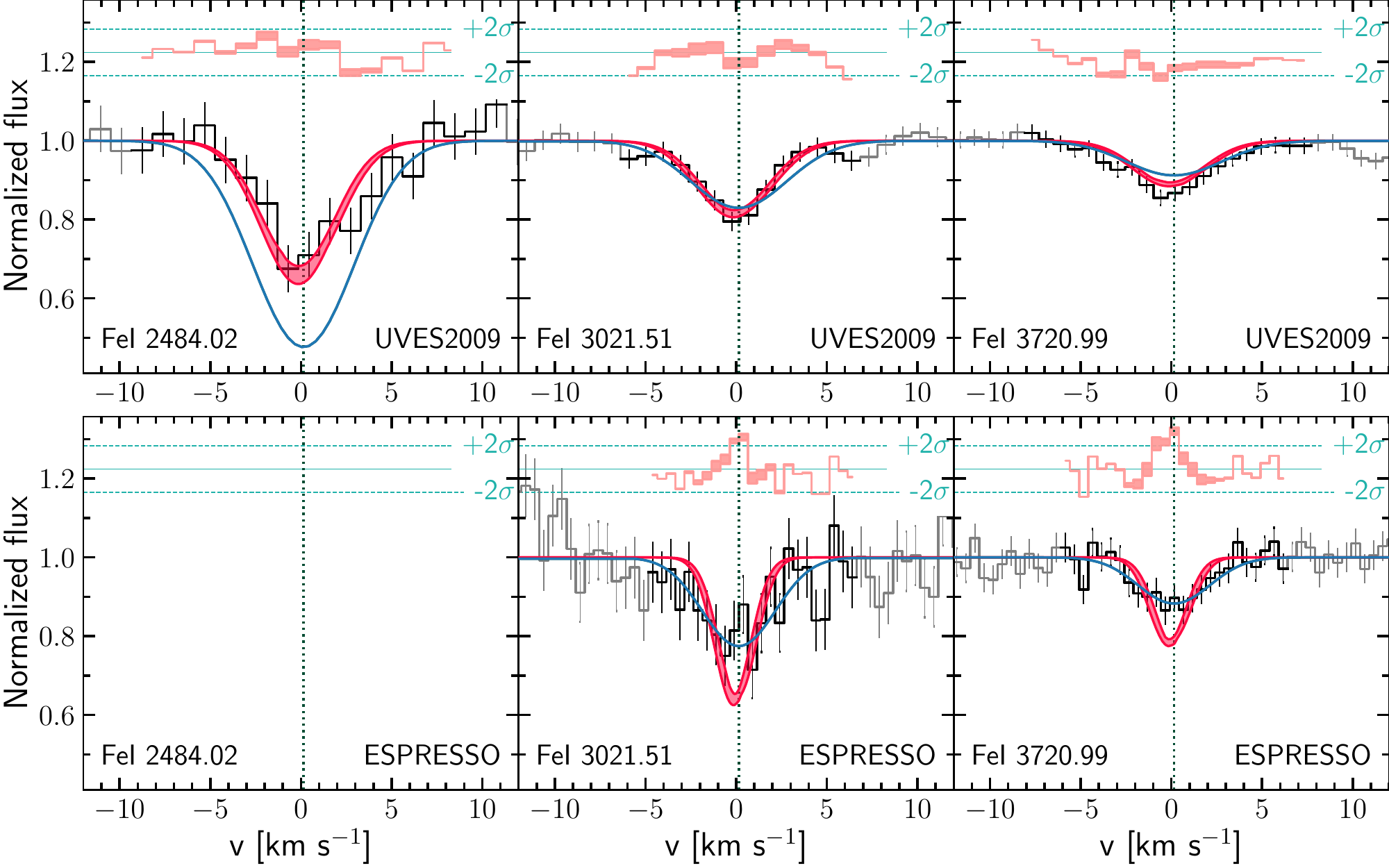}
    \caption{Fit to the \FeI\ absorption lines with one-component model. From top to bottom, the black line corresponds to the data obtained with UVES 2009 and ESPRESSO 2018 (\FeI$\lambda2484$\AA\ line is missed in ESPRESSO spectrum), the red and blue profiles are the one-component models constrained respectively from UVES and ESPRESSO data only. } 
    \label{f:FeI_compar}
\end{figure}

\begin{figure*}
    \centering
    \includegraphics[clip=,width=\hsize]{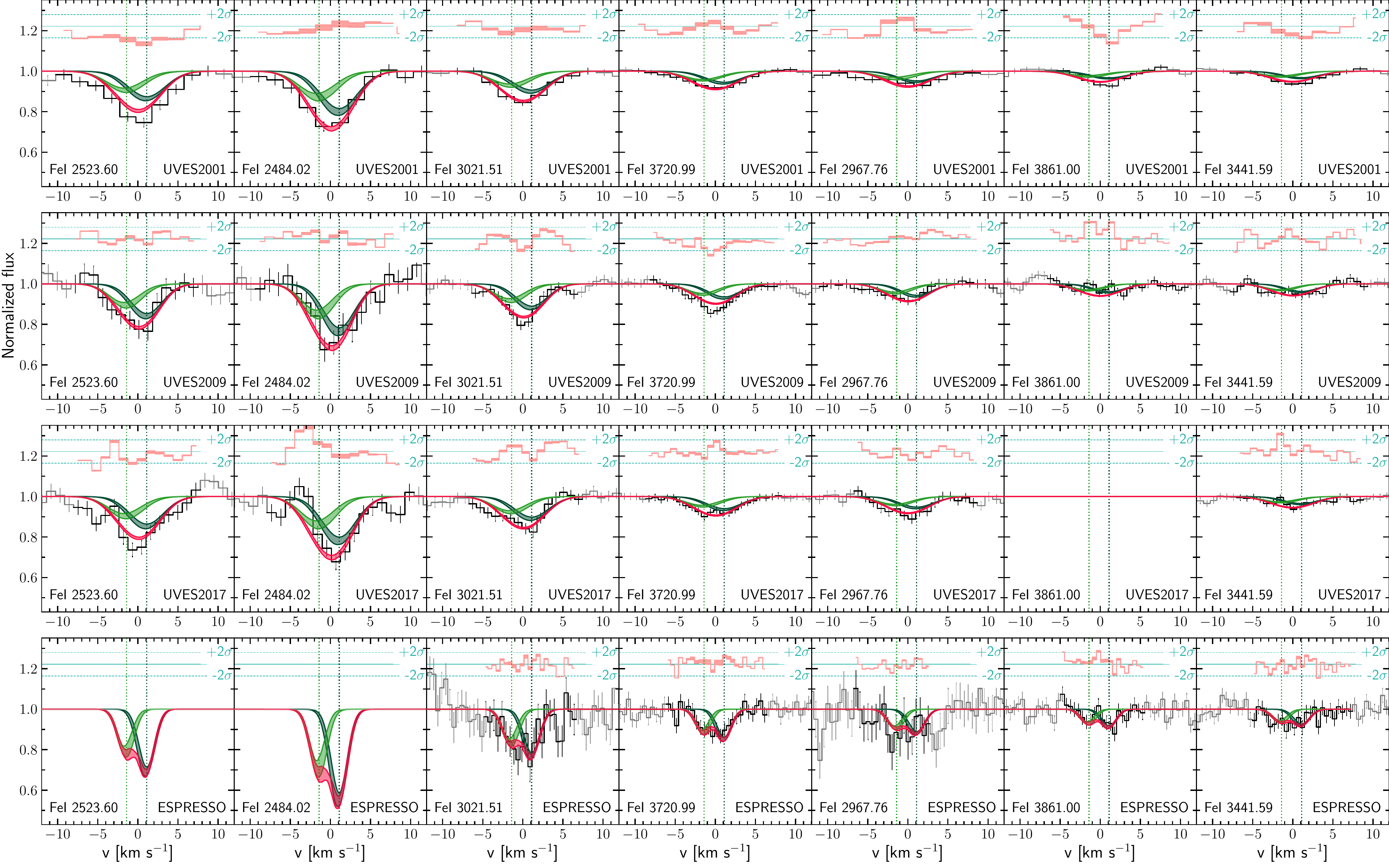}
    \caption{Fit to the \FeI\ absorption lines with a two component model. From top to bottom, the black line corresponds to the data obtained with UVES (2001, 2009 and 2017) and ESPRESSO (2018). The fit profiles are represented using 0.68 central percentile interval of their own distributions sampled from the obtained posterior probability distribution of the fitting parameters. Red and green areas correspond to the total profile and contribution of each component, respectively.
    Residuals are shown above each panel. 
    } 
    \label{f:FeI_2comp}
\end{figure*}

{Following previous UVES studies, we first fitted the \ion{Fe}{I} lines with a single component Voigt-profile.}
{We used combined spectra obtained for each set of observations. In other words, we used four different spectra: three from UVES taken in 2001, 2009 and 2017 and our new ESPRESSO spectrum. To fit the absorption lines for each spectra we adopted the average resolution powers of 55000, 65000, 60000 for 2001, 2009 and 2017 UVES spectra, respectively. For the ESPRESSO spectrum, we used different resolution for the blue and red channels, as provided in Table~\ref{tab:obslog}.} However, we could not obtain a simultaneously satisfactory fit of the {ESPRESSO + UVES} data taken altogether.  
Considering each spectrum independently, we found differences in the derived \FeI\ parameters between epochs, the most noticeable difference being in the Doppler parameter, which we found to be significantly higher in the ESPRESSO spectrum ($2.4^{+0.5}_{-0.2}$\,\kms) than suggested by any of the UVES spectra ($\sim 0.6$\,\kms) with typical uncertainty less than $0.1$\,\kms.\footnote{We note that these Doppler parameters are much lower than the UVES resolution element, but they are constrained from the relative strengths of the \FeI\ lines, since several lines with different oscillator strength were used.} The inconsistency between {the one-component models fitted to} the ESPRESSO and UVES data {independently} is shown in Fig.~\ref{f:FeI_compar}.
While it is tempting to attribute this to time variation of the line profiles, such an interpretation is suspicious since we did not see any significant changes between UVES 2001 and 2017, and the difference appears only in the new ESPRESSO spectrum. Therefore it is more likely that the difference is actually due to an inappropriate model of the velocity profile. {Indeed, one can note in Fig.~\ref{f:FeI_compar} that the observed \FeI$\lambda3021$\AA\ and 3720\AA\ line profiles for ESPRESSO and UVES have about the same width when the resolution of the two instruments differs by a factor of 2.5. This strongly suggests the presence of more than one component.} 

Fitting the \FeI\ lines with a two-component model solves the issue (see Fig.~\ref{f:FeI_2comp}). The whole set of UVES and ESPRESSO data is now well fitted with the same set of parameters, which is given in Table~\ref{tab:FeI_results}. 
Remarkably, we found Doppler parameters of only $b\sim 0.3$~\kms. Assuming thermal broadening only, this corresponds to a kinetic temperature $T\sim 300$~K. Since turbulent motions can also contribute to the line broadening, the actual temperature 
must be lower than this value, in agreement with the $T\sim 100$~K derived from modelling the physical conditions in the gas. 

\begin{figure}
    \centering
    \includegraphics[clip=,width=1.0\hsize]{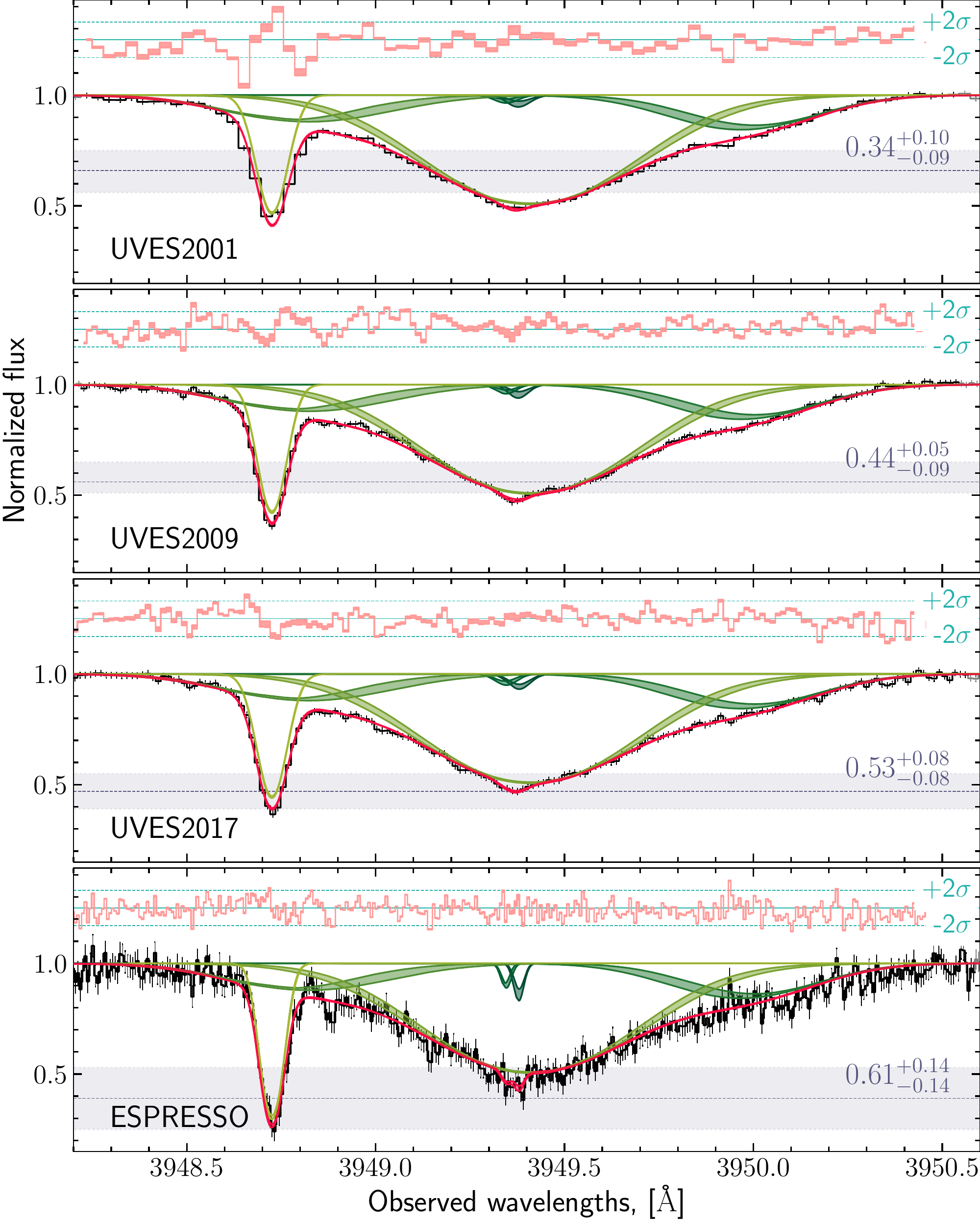}
    \caption{Fit to the the region around $\FeI\lambda2719\,$\AA\ line at $z=0.452$. From top to bottom, the black line corresponds to the data obtained with UVES 2001, 2009, 2017 and ESPRESSO 2018. The fit profiles are presented in the same manner as in Fig.~\ref{f:FeI_2comp}. The horizontal dashed line and grey stripe show $1 - C_f$ and its uncertainty, respectively. The constrained value of $C_f$ for each spectrum is given in the bottom-right of each panel. 
    } 
    \label{f:FeI_2719}
\end{figure}

{To take into account the possible partial coverage in \FeI$\lambda2719$\AA\ line, we considered a modified line profile with simple partial covering model \citep[see e.g.][]{Ganguly1999}, where the covering factor is an independent and free parameter during the fit.}
We confirm the partial coverage of the \lya\ emission by the cold, \FeI-bearing gas with $C_f\approx0.4-0.6$, assuming the same covering factor for both \FeI\ components and for all exposures. Alternatively, fitting the spectra with independent covering factor for each exposure, we obtained a gradual increase of the covering factor, $C_f$, with time from $0.34^{+0.10}_{-0.09}$ in 2001 (consistent with values obtained by \citealt{Bergeron2017}) to $0.61^{+0.14}_{-0.14}$ in 2018 (see Fig.~\ref{f:FeI_2719}), although this is not highly statistically significant given the relatively high uncertainties on individual values. 

The continuum emission accounts for about 30\% of the total observed flux at the position of \lya\, and is more likely fully covered by the \FeI-absorbing cloud, since we do not see any evidence for partial coverage at other wavelengths. This is also expected since the accretion disc responsible for the UV continuum emission is expected to have a size of at most $\sim10^{-3}$ pc.
The time-variation of $C_f$, if confirmed, would suggest that the covering fraction of the \lya\ emission line flux increases from $0.05^{+0.14}_{-0.05}$ to $0.47^{+0.19}_{-0.19}$ between 2001 and 2018 (see Eq. 4 of \citealt{Bergeron2017}).
Since the ratio of the continuum to \Lya \ emission line flux does not vary significantly (less than 20\% over the same period), this indicates that the possible time variation of $C_f$ is likely attributed to the relative motion of the \FeI-bearing cloud with respect to the line of sight. It also means that 
a large fraction of \lya\ photons arise from a region smaller in projection than the \FeI-bearing cloud.
In other words, at least half of the \lya\ photons with velocities corresponding to the position of the \FeI$\lambda2719$\,\AA\ line is produced in the very nuclear region of the quasar. It would be interesting to continue monitoring this system to see 
if the covering fraction continues to increase, setting an upper limit to the amount of photons arising from (or scattered by)  
gas in more extended regions around the quasar nucleus. 

This work shows that resolving the velocity structure along the line-of-sight is a crucial step towards a physical interpretation of the data. This includes the investigation of possible variations of the line profiles (e.g. covering factors) due to the transverse motion of the {cloud with respect to} the line of sight. 

\begin{table}[]
\renewcommand{\arraystretch}{1.3}
\setlength{\tabcolsep}{3pt}
    \centering
        \caption{Results from fitting UVES and ESPRESSO altogether}
        
    \begin{tabular}{c c c c c}
    \hline\hline
    Species & $z$                         & $b$                    & $\log N(\rm X)$             & $C_f$\tablefootmark{a} \\
     (X)    &                             & (\kms)                 &                         &       \\
     \hline
     \FeI   & $0.4520546(^{+5}_{-4})$ & $0.22^{+0.05}_{-0.03}$ & $11.94^{+0.03}_{-0.06}$ & \multirow{2}{*}{$0.45^{+0.06}_{-0.05}$} \\
     \FeI   & $0.4520668(^{+6}_{-5})$     & $0.39^{+0.06}_{-0.04}$ & $12.02^{+0.03}_{-0.04}$ & \\
     \hline
     \HI    & $2.248242(^{+9}_{-12})$    & $19.9^{+1.3}_{-1.0}$   & $12.50^{+0.03}_{-0.03}$ & \multirow{4}{*}{1} \\
     \HI    & $2.2487523(^{+20}_{-20})$   & $27.4^{+0.7}_{-0.5}$   & $13.39^{+0.01}_{-0.01}$ &  \\
     \HI    & $2.249235(^{+4}_{-11})$      & $18.8^{+0.7}_{-0.5}$   & $12.58^{+0.03}_{-0.03}$ &  \\
     \SiII  & $1.5864332(^{+1}_{-3})$  & $2.65^{+0.07}_{-0.07}$ & $13.07^{+0.01}_{-0.01}$ &  \\
    \hline
    \end{tabular}
    \label{tab:FeI_results}
    \renewcommand{\arraystretch}{1.}
    \tablefoot{
    \tablefoottext{a}{The covering fraction correspond to that at the wavelength of the \lya\ emission line. It is 
    less than unity for \FeI$\lambda2719$ only. Other \FeI\ lines are redshifted at wavelengths corresponding to the quasar continuum emission only.} 
    }
\end{table}


\section{Summary \label{s:concl}}

We have presented the analysis of a very high-resolution spectrum ($R \gtrsim 130\,000$) of the quasar \qso\ obtained
using the new VLT/ESPRESSO spectrograph. Our observations cover the range $380-788$~nm with exquisite
spectral fidelity and wavelength-calibration accuracy. These characteristics allow us to investigate the velocity
profiles of several absorption systems with unprecedented precision.

We have disentangled the thermal and turbulent broadening in individual components of the metal absorption-line complex (spread over about 400~\kms) associated with a low-metallicity sub-DLA
at $z=2.187$. We infer temperatures of $\sim16\,000$\,K, i.e., twice the canonical Galactic WNM value. By compiling temperature measurements in other low-metallicity high-$z$ systems, we reveal an anti-correlation between the kinetic temperature in individual velocity components and the total \HI\ column density integrated over the profile.
Interestingly, the systems have metallicities, \HI\ column densities, and temperatures in-between those found in
the warm neutral ISM of galaxy discs and those seen in the intergalactic medium. We are then possibly
witnessing gas cooling in the circum-galactic medium, in a transition region between IGM and ISM.  This finding opens up exciting prospects to further investigate the thermal state of the gas around galaxies using high-resolution spectroscopy.

The high spectral resolution and wavelength accuracy allowed us to constrain the Mg isotopic ratio 
at $z=0.45$ and $z=1.65$, which we find to be consistent with the Solar value. Apparent strong enhancement of
heavy Mg isotopes previously reported in the literature for these systems were hence likely due to
distortion in the UVES wavelength scale.
This also shows that ESPRESSO is well suited for such isotopic measurements in various absorption
systems, opening a new way to place constraints on the stellar nucleosynthetic history over cosmic time
(see also \citealt{Welsh2020}).

Finally, we confirm partial coverage of the background source at $\lya$ (quasar rest-frame) wavelengths by the
\FeI\ absorption system at $\zabs=0.45$. In other words, the background source cannot be considered as an infinitely narrow beam anymore; About half of the quasar photons at these wavelengths do not pass through
the absorber.
Thanks to the combined analysis of UVES and ESPRESSO data, we were also able to reveal velocity sub-structure
within the \FeI-bearing gas, that presents two components with Doppler parameters of only $b\sim 0.3$~\kms,
confirming the low kinetic temperature obtained from modelling of the gas physical conditions.

Our results are mostly S/N-limited so that more stringent constraints can be obtained by the addition of further exposures. Our work highlights several important science cases that can be addressed
from single-object high-resolution spectroscopy with large telescopes. These will certainly become routinely
addressed with future ground-based instruments such as HIRES on the E-ELT {\citep{Marconi2021}}, G-CLEF on
the GMT \citep{Gclef2016}, and HROS on the TMT \citep{Hros2006}, as well as Pollux on the LUVOIR space telescope \citep{Bouret2018}.

\begin{acknowledgements}

{We thank the anonymous referee for detailed and constructive comments.}
We thank Michael Murphy for discussions on the \MgII\ wavelengths and isotopic-ratio
measurements, and Peter Shternin for discussions on Bayesian hierarchical modelling. We
acknowledge support from the French {\sl Agence Nationale de la Recherche} under ANR
grant 17-CE31-0011-01 / project ``HIH2'' (PI Noterdaeme). SB and KT are partially supported by RFRB grant
18-52-15021 and the Basis Foundation. CL thanks ESO for support during a Research Period.
GD thanks IAP for hospitality when this work was initiated. SL was funded by project FONDECYT 1191232.

\end{acknowledgements}

\bibliographystyle{aa}
\bibliography{HE0001}

\appendix

\input{continuum.tex}

\begin{figure*}
    \centering
    \includegraphics[clip=, width=0.9\hsize]{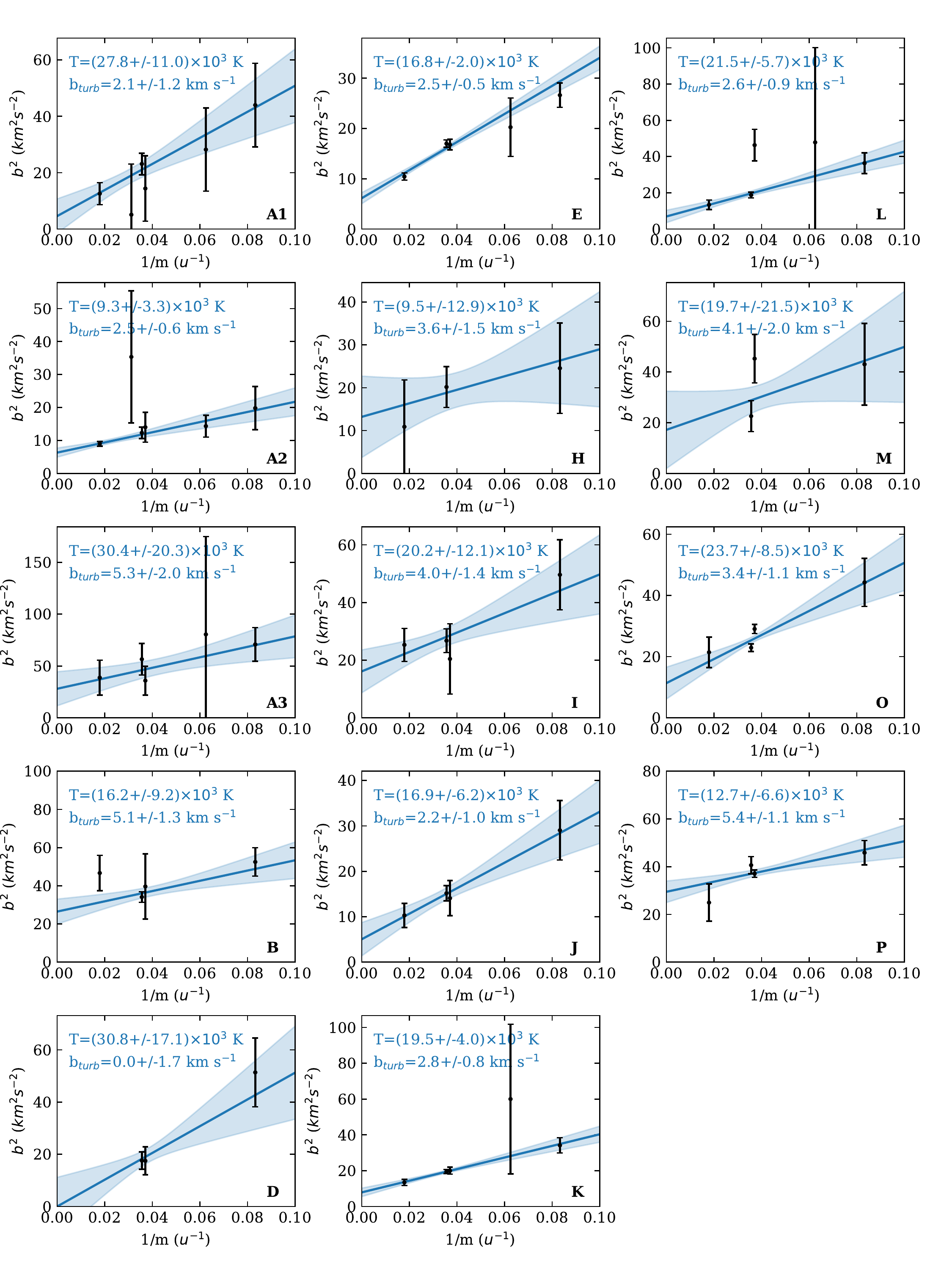} 
    \caption{{Temperature and turbulent broadening parameter determinations
    using independent Doppler parameters for each species (i.e., method 2) in various components. For the weakest component (C, G and N), the uncertainties are too large for providing meaningful constraints.  
    Black points and error bars represent the square of the
    total Doppler parameter $b$ measured for various species (\FeII, \SII\, \SiII, \AlII, \OI\ and \CII) as a function of the inverse of their mass.
    The blue line and shaded area represent the linear fit to the data and 1\,$\sigma$ confidence band.}}  
    \label{f:T_by_meth_2}
\end{figure*}

\end{document}

%% file: tab_obslog.tex
\begin{table*}
\centering
\caption{Observational data.\label{tab:obslog}}
\begin{tabular}{rllllccccr}
\hline\hline
{\large\strut}
Date range & PI & VLT & Instrument & Setup & Slit\tablefootmark{a} & Binning & $T_{\rm tot}$\tablefootmark{b} & IQ\tablefootmark{c} & $R$\tablefootmark{d} \\
           &    &     &            &       & [$\arcsec$]           &         & [s]                            & [$\arcsec$]         & [$\times 10^3$]      \\
\hline
{\large\strut}
09-28/11/2018 & Ledoux   & UT3 & ESPRESSO & SingleUT-HR & ... & 2$\times$1 & 24\,110 & 0.89 & 132.0/146.0    \\
\hline
{\large\strut}
21-26/08/2017 & Bergeron & UT2 & UVES     & B390+R564   & 0.8 & 1$\times$1 & 21\,000 & 0.83 &  58.8/59.4/58.3\\
21-24/09/2009 & Molaro   & UT2 & UVES     & B390+R580   & 0.7 & 1$\times$1 & 33\,870 & 0.92 &  64.8/73.4/65.5\\
20/09/2009    & Molaro   & UT2 & UVES     & B437+R760   & 0.7 & 1$\times$1 & 16\,200 & 0.78 &  64.9/72.5/67.3\\
22/09/2009    & Molaro   & UT2 & UVES     & (B420+)R700 & 0.7 & 1$\times$1 & 15\,000 & 1.18 & (...)/72.5/67.3\\
29/06/2001    & Bergeron & UT2 & UVES     & B437+R860   & 1.0 & 2$\times$2 &  3\,600 & 0.53 &  60.3/57.3/56.1\\
15-24/08/2001 & Bergeron & UT2 & UVES     & B437+R860   & 1.0 & 2$\times$2 & 18\,000 & 0.87 &  48.1/48.7/45.7\\
16-24/08/2001 & Bergeron & UT2 & UVES     & B346+R580   & 1.0 & 2$\times$2 & 21\,600 & 0.55 &  58.6/66.3/59.3\\

\hline
\end{tabular}
\tablefoot{\tablefoottext{a}{Slit width.}
\tablefoottext{b}{Total on-source exposure time.}
\tablefoottext{c}{Image Quality from the Shack-Hartmann sensor of the VLT active optics
at 650~nm at the airmass of observations.}
\tablefoottext{d}{Resolving power measured from emission lines in ThAr calibration frames. The
listed values correspond to the Blue and Red arm spectra, respectively, for ESPRESSO, and to the
Blue, Lower-Red, and Upper-Red spectra for UVES. For the latter instrument, the effect of an IQ
smaller than the slit width in some cases is taken into account.}
}
\end{table*}

%% file: systems.tex
\begin{table}[]
    \centering
    \caption{Absorption lines identified in the ESPRESSO spectrum of \qso.}
\begin{tabular}{c c l l }
\hline\hline
{\large \strut}$z_{\rm abs}$\tablefootmark{a}   & $v_{\rm range}$\tablefootmark{b}  & species & line IDs \\
               & (\kms)          &         & ({\AA}) \\
\hline
0.00001  &               &  \NaI & 5891, 5897         \\ 
                \noalign{\smallskip}
0.27052  &               &  \CaII & 3934, 3969        \\ %
                \noalign{\smallskip}
\textbf{0.45206}  & [0,+150]      &  \FeI &  2719, 2967, 3021, 3441  \\ %
         &               &       &  3720, 3825, 3861  \\ %
         &               &  \MgII & 2796, 2083        \\ %
         &               &  \MgI  & 2852              \\ %
         &               &  \CaII & 3934, 3969        \\ %
         &               &  \CaI  & 4227              \\ %
                \noalign{\smallskip}
0.94902  & [-60,+130]    &  \MgII & 2796, 2803        \\ %
         &               &  \MgI  & 2852              \\ %
         &               &  \FeII & 2586,2600,2344, 2382    \\ %
                \noalign{\smallskip}
1.58643  & [-200,+150]   &  \CIV  & 1548, 1550        \\ %
         &               &  \MgII & 2796, 2803        \\ %
         &               &  \MgI  & 2852              \\ %
         &               &  \SiII & 1526              \\ %
         &               &  \FeII & 1608, 2344, 2374, 2382  \\ %
         &               &        & 2586, 2600  \\ %
                \noalign{\smallskip}
\textbf{1.65147}  & [-40,+40]    &  \CIV  & 1548, 1550        \\ %
         &               &  \SiII & 1526              \\ %
         &               &  \AlIII& 1854 ,1862        \\ %
         &               &  \MgII & 2796, 2803        \\ %
         &               &  \MgI  & 2852              \\ %
                \noalign{\smallskip}
1.66602  & [-100,+120]   &  \CIV  & 1548, 1550        \\ %
                \noalign{\smallskip}
1.72721  &               &  \CIV  & 1548, 1550        \\ %
                \noalign{\smallskip}
1.82737  & [-40,+50]     &  \CIV  & 1548, 1550        \\ %
         &               &  \SiIV & 1393, 1402        \\ %
                \noalign{\smallskip}
2.03228  & [-50,+50]     &  \CIV  & 1548, 1550        \\ %
                \noalign{\smallskip}
\textbf{2.18716}  & [-450,+30]    &  \HI   & 1215              \\ %
         &               &  \DI   & 1215              \\ %
         &               &  \CIV  & 1548, 1550        \\ %
         &               &  \SiII & 1260, 1304, 1526  \\ %
         &               &  \FeII & 1260, 1608, 2344, 2374  \\ %
         &               &        & 2382        \\ %
         &               &  \OI   & 1302              \\ %
         &               &  \CII  & 1334              \\ %
         &               &  \SiIV & 1392, 1402        \\ %
         &               &  \AlII & 1670              \\ %
         &               &  \AlIII& 1854, 1682        \\ %
                         \noalign{\smallskip}
2.25710  &               &  \CIV  & 1548, 1550        \\ %
\hline
    \end{tabular}
     \tablefoot{
 \tablefoottext{a}{Redshifts marked in bold face correspond to systems considered in this work.} 
 \tablefoottext{b}{In case of an absorption complex spread over more than a few tens of \kms, we provide in this column the velocity range over which components are detected (with respect to the main absorption redshift given in the first column).}
 }
    \label{t:systems}
\end{table}

%% file: DLAmet_newfwhm.tex
\longtab{
\begin{longtable}{llllllll}
\caption{Results from Voigt-profile fitting of metal species associated to the DLA at $z=2.187$.\label{t:DLAmet}
}\\
\hline
\hline
{\Large \strut} Comp & $v_{\rm rel}$  (\kms) & \zabs\ & Ion & $\log N$   (\cmsq)  & $b_{\rm turb}$  (\kms) & $T (10^{4}$\,K) & $b_{\rm tot}$  \\
\hline
\endfirsthead

\caption{continued.}\\
\hline\hline
{\Large \strut} Comp & $v_{\rm rel}$  (\kms) & \zabs\ & Ion & $\log N$   (\cmsq)  & $b_{\rm turb}$  (\kms) & $T (10^{4}$\,K) & $b_{\rm tot}$  \\
\hline
\endhead

A1  &  +24 & 2.18725 & \CII   & 12.96 $\pm$ 0.08  & 1.96 $\pm$ 1.35 & 2.99 $\pm$ 0.99 & 6.72 \\
    &      &         & \OI    & 13.16 $\pm$ 0.08  &                 &                 & 5.91 \\
    &      &         & \AlII  & 10.73 $\pm$ 0.16  &                 &                 & 4.72 \\
    &      &         & \SiII  & 12.22 $\pm$ 0.04  &                 &                 & 4.64 \\
    &      &         & \SII   & 12.98 $\pm$ 0.19  &                 &                 & 4.40 \\
    &      &         & \FeII  & 11.81 $\pm$ 0.05  &                 &                 & 3.57 \smallskip\\
    
A2  &  +15 & 2.18716 & \CII   & 13.86 $\pm$ 0.09  & 2.44 $\pm$ 0.18 & 0.99 $\pm$ 0.25 & 4.43 \\
    &      &         & \OI    & 14.48 $\pm$ 0.08  &                 &                 & 4.03 \\
    &      &         & \AlII  & 11.52 $\pm$ 0.09  &                 &                 & 3.47 \\
    &      &         & \SiII  & 13.11 $\pm$ 0.03  &                 &                 & 3.44 \\
    &      &         & \SII   & 13.16 $\pm$ 0.08  &                 &                 & 3.33 \\
    &      &         & \FeII  & 12.79 $\pm$ 0.02  &                 &                 & 2.98 \smallskip\\
    
A3  &  +11 & 2.18711 & \CII   & 13.62 $\pm$ 0.07  & 5.45 $\pm$ 0.76 & 2.94 $\pm$ 0.95 & 8.39 \\
    &      &         & \OI    & 13.06 $\pm$ 0.24  &                 &                 & 7.76 \\
    &      &         & \AlII  & 11.74 $\pm$ 0.06  &                 &                 & 6.91 \\
    &      &         & \SiII  & 12.71 $\pm$ 0.07  &                 &                 & 6.86 \\
    &      &         & \FeII  & 12.17 $\pm$ 0.08  &                 &                 & 6.20 \smallskip\\
    
B   &  -29 & 2.18669 & \CII   & 13.18 $\pm$ 0.03  & 5.46 $\pm$ 0.55 & 1.28 $\pm$ 0.82 & 6.90 \\
    &      &         & \AlII  & 11.00 $\pm$ 0.06  &                 &                 & 6.14 \\
    &      &         & \SiII  & 12.33 $\pm$ 0.02  &                 &                 & 6.12 \\
    &      &         & \FeII  & 11.86 $\pm$ 0.03  &                 &                 & 5.80 \smallskip\\
    
C   & -111 & 2.18583 & \CII   & 12.51 $\pm$ 0.10  & 5.94 $\pm$ 2.35 & $<$ 3.01 & 5.94 \\
    &      &         & \SiII  & 11.54 $\pm$ 0.07  &                 &                 & 5.94 \\
    &      &         & \AlII  & 10.73 $\pm$ 0.12  &                 &                 & 5.94 \smallskip\\

D   & -145 & 2.18546 & \CII   & 12.99 $\pm$ 0.05  & \ldots & 3.23 $\pm$ 1.52 & 6.69 \\
    &      &         & \AlII  & 11.17 $\pm$ 0.04  &                 &                 & 4.46 \\
    &      &         & \SiII  & 11.87 $\pm$ 0.03  &                 &                 & 4.37 \smallskip\\

E   & -160 & 2.18530 & \CII   & 13.61 $\pm$ 0.02  & 2.46 $\pm$ 0.24 & 1.71 $\pm$ 0.22 & 5.45 \\
    &      &         & \OI    & 13.21 $\pm$ 0.05  &                 &                 & 4.88 \\
    &      &         & \AlII  & 11.87 $\pm$ 0.01  &                 &                 & 4.07 \\
    &      &         & \SiII  & 12.85 $\pm$ 0.01  &                 &                 & 4.02 \\
    &      &         & \FeII  & 12.29 $\pm$ 0.01  &                 &                 & 3.34 \smallskip\\

G   & -226 & 2.18460 & \CII   & 13.15 $\pm$ 0.09  & 10.44 $\pm$ 4.18 & $<$ 21.16 & 15.98 \\
    &      &         & \AlII  & 11.35 $\pm$ 0.07  &                 &                 & 13.19 \\
    &      &         & \SiII  & 12.12 $\pm$ 0.08  &                 &                 & 13.10 \smallskip\\

H   & -230 & 2.18455 & \CII   & 12.53 $\pm$ 0.20  & 2.44 $\pm$ 1.73 & $<$ 2.02 & 3.88 \\
    &      &         & \SiII  & 11.81 $\pm$ 0.07  &                 &                 & 3.14 \\
    &      &         & \FeII  & 11.09 $\pm$ 0.14  &                 &                 & 2.81 \smallskip\\

I   & -240 & 2.18445 & \CII   & 12.95 $\pm$ 0.10  & 4.31 $\pm$ 0.83 & 1.45 $\pm$ 1.33 & 6.22 \\
    &      &         & \AlII  & 10.98 $\pm$ 0.12  &                 &                 & 5.24 \\
    &      &         & \SiII  & 12.11 $\pm$ 0.06  &                 &                 & 5.21 \\
    &      &         & \FeII  & 11.82 $\pm$ 0.04  &                 &                 & 4.78 \smallskip\\

J   & -269 & 2.18414 & \CII   & 13.04 $\pm$ 0.04  & 2.27 $\pm$ 0.80 & 1.68 $\pm$ 0.61 & 5.33 \\
    &      &         & \AlII  & 11.20 $\pm$ 0.04  &                 &                 & 3.94 \\
    &      &         & \SiII  & 12.12 $\pm$ 0.02  &                 &                 & 3.88 \\
    &      &         & \FeII  & 11.65 $\pm$ 0.04  &                 &                 & 3.18 \smallskip\\

K   & -282 & 2.18400 & \CII   & 13.40 $\pm$ 0.02  & 2.79 $\pm$ 0.43 & 1.99 $\pm$ 0.41 & 5.94 \\
    &      &         & \OI    & 12.82 $\pm$ 0.10  &                 &                 & 5.33 \\
    &      &         & \AlII  & 11.68 $\pm$ 0.02  &                 &                 & 4.47 \\
    &      &         & \SiII  & 12.57 $\pm$ 0.01  &                 &                 & 4.42 \\
    &      &         & \FeII  & 11.99 $\pm$ 0.02  &                 &                 & 3.70 \smallskip\\

L   & -305 & 2.18376 & \CII   & 13.20 $\pm$ 0.03  & 2.84 $\pm$ 0.66 & 2.19 $\pm$ 0.61 & 6.20 \\
    &      &         & \OI    & 12.61 $\pm$ 0.16  &                 &                 & 5.55 \\
    &      &         & \AlII  & 11.36 $\pm$ 0.03  &                 &                 & 4.64 \\
    &      &         & \SiII  & 12.29 $\pm$ 0.01  &                 &                 & 4.59 \\
    &      &         & \FeII  & 11.78 $\pm$ 0.03  &                 &                 & 3.82 \smallskip\\

M   & -327 & 2.18352 & \CII   & 12.96 $\pm$ 0.07  & 6.30 $\pm$ 1.40 & $<$ 3.70  & 7.50 \\
    &      &         & \AlII  & 11.55 $\pm$ 0.03  &                 &                 & 6.86 \\
    &      &         & \SiII  & 12.03 $\pm$ 0.04  &                 &                 & 6.84 \smallskip\\

N   & -339 & 2.18340 & \CII   & 12.57 $\pm$ 0.17  & 4.01 $\pm$ 3.23 & $<$ 4.03 & 5.35 \\
    &      &         & \AlII  & 10.79 $\pm$ 0.17  &                 &                 & 4.65 \\
    &      &         & \SiII  & 11.46 $\pm$ 0.13  &                 &                 & 4.63 \smallskip\\

O   & -355 & 2.18323 & \CII   & 13.48 $\pm$ 0.03  & 3.68 $\pm$ 0.65 & 2.22 $\pm$ 0.79 & 6.66 \\
    &      &         & \OI    & 12.40 $\pm$ 0.25  &                 &                 & 6.06 \\
    &      &         & \AlII  & 12.20 $\pm$ 0.01  &                 &                 & 5.22 \\
    &      &         & \SiII  & 12.98 $\pm$ 0.01  &                 &                 & 5.17 \\
    &      &         & \FeII  & 11.74 $\pm$ 0.04  &                 &                 & 4.49 \smallskip\\

P   & -369 & 2.18308 & \CII   & 13.57 $\pm$ 0.02  & 5.47 $\pm$ 0.42 & 1.25 $\pm$ 0.68 & 6.88 \\
    &      &         & \OI    & 12.34 $\pm$ 0.30  &                 &                 & 6.55 \\
    &      &         & \AlII  & 12.37 $\pm$ 0.01  &                 &                 & 6.14 \\
    &      &         & \SiII  & 13.07 $\pm$ 0.01  &                 &                 & 6.11 \\
    &      &         & \FeII  & 11.65 $\pm$ 0.05  &                 &                 & 5.80 \smallskip\\

\hline
\end{longtable}
}

%% file: continuum.tex
\section{
Hierarchical Bayesian modelling of the Fe\,{\sc i} absorber}
\label{s:feifit}

We describe here the method used to fit the \ion{Fe}{i} absorption lines with Voigt profiles in Section~\ref{s:PC}. Since most of the \ion{Fe}{I} lines are weak, we were forced to take into account continuum placement uncertainties. To do it rigorously, we used a hierarchical Bayesian approach.   
In a standard Bayesian approach, the distribution of the model parameters is given by
\begin{equation}
    p(\theta|D) \propto  p(D|\theta) p(\theta),
\end{equation}
where $p(\theta)$ is a prior on the model parameters (denoted by $\theta$) and $D$ stands for observed data. The likelihood function $p(D|\theta)$, under the assumption of Gaussian distribution of the pixel uncertainties can be written as
\begin{equation}
    \label{eq:lnL_regular}
    p(D|\theta) \propto \prod\limits_{i}\exp\left[-\frac{1}{2}\left(\frac{m(x_i) - y_i}{\sigma_i}\right)^2\right],
\end{equation}
where $x_i$, $y_i$ and $\sigma_i$ are the wavelength, flux and uncertainty in the $i^{\rm th}$ pixel of the spectrum, respectively. Evidently $D \equiv (x_i, y_i, \sigma_i)$. The model profile, $m$, is usually specified for absorption lines as
\begin{equation}
    \label{eq:m_regular}
    m(x) = \int C(x') e^{-\tau(x')} I(x,x') dx',
\end{equation}
where $\tau(x)$ is an optical depth describing the profiles of absorption lines, $I(x,x')$ is the instrument function, and $C(x)$ denotes the unabsorbed continuum of the background source. When the background source is a quasar, 
the continuum does not have a simple shape. In addition, the spectra are generally not flux-calibrated, and the presence of absorption lines unrelated to the absorption system in question affects the apparent continuum. 
Therefore, the unabsorbed quasar continuum is usually determined by interpolating the observed spectrum using regions unaffected by absorption, as determined by visual inspection. 
B-spline functions are frequently used to this end. This works well for relatively strong absorption lines and does not introduce significant biases on the inferred absorption model parameters. However, for weak absorption lines, the subjectivity of continuum placement can have stronger consequences. 

To take into account this uncertainty, we propose the following extension of the Bayesian model. 
For simplicity, let us describe first a single absorption line.
The observed spectrum is subject to random statistical noise, resulting in the dispersion of spectral pixels, around some smooth envelope. In the model situation, this dispersion corresponds to the value of the pixel uncertainties, i.e. to the spectral quality. Hence the dispersion of the continuum placement (the difference between the constructed and true continuum) is expected to be proportional to $\bar{\sigma}$ -- a mean pixel uncertainty within some wavelength range in the vicinity of the line. 
Therefore it is natural to describe the continuum displacement in the units of $\bar{\sigma}$. This is particularly useful for the modelling of multiple absorption lines in distinct wavelength ranges, with different spectral quality.
In the first approximation, the effect of continuum placement uncertainty can be addressed by changing the continuum following
\begin{equation}
    C(x) \to C(x) (1 + \bar{\sigma} \alpha),
\end{equation}
where  
$\alpha$ is a parameter specifying the continuum displacement in the units of $\bar{\sigma}$. Therefore a model profile $\tilde m(x)$ that takes into account continuum uncertainty can be expressed as
\begin{equation}
    \label{eq:m_correction}
    \tilde m(x) = m(x) (1 + \bar{\sigma} \alpha).
\end{equation}

One can in principle use $\alpha$ as a nuisance parameter during the fit and constrain absorption model parameters jointly with $\alpha$. 
The result, however, is that an improper choice of the continuum model may bias the absorption model parameters. Additionally, in case of multiple absorption lines, each line (or line region, if some lines are blended) should have its own $\alpha$. This may lead to an undesired significant increase of the number of model parameters. This can be circumvented by using a hierarchical Bayesian approach.

We assume that $\alpha$ follows some distribution $p(\alpha|d)$ described by some parameter $d$. This corresponds to the sampling of the continuum placement, i.e. we mimic the situation when different people obtain independent continuum placement. The posterior distribution of the parameters can then be written as
\begin{equation}
    \label{eq:lnL_modified}
    p(\theta,\alpha,d|D) \propto p(D|\theta,\alpha) p(\alpha|d)  p(d) p(\theta),
\end{equation}
where $\theta$ is the set of the model parameters describing only the absorption profile, and $p(d)$ is a prior on $d$. Since we are not interested in the distribution of $\alpha$, we can marginalise over this parameter
\begin{equation}
    \label{eq:lnL_marginalize}
    p(\theta,d|D) \propto \left(\int p(D|\theta,\alpha) p(\alpha|d)  d\alpha \right) p(d) p(\theta),
\end{equation}
which gives that the integral in the bracket is a modified likelihood function, $p(D|\theta, d)$. 

Considering that $p(D|\theta,\alpha)$ is described by equation~\eqref{eq:lnL_regular} with replacement \eqref{eq:m_correction}, and assuming that $p(\alpha|d)$ is a Gaussian function 
with zero mean and dispersion $d$, we can take an integral over $\alpha$ in analytical form, which gives
\begin{equation}
    p(D|\theta, d) = p(D|\theta) \cdot \frac1{\sqrt{2\pi A}d} \exp\left(\frac{B^2}{2A}\right),
\end{equation}
where
\begin{align}
    \label{eq:A}
    A &= \sum\limits_i \frac{m(x_i)^2\bar{\sigma}_i^2}{\sigma_i^2} + \frac1{d^2} \\
    \label{eq:B}
    B &= \sum\limits_i \frac{m(x_i)\bar{\sigma}_i(y_i-m(x_i))}{\sigma_i^2},
\end{align}
where $\bar{\sigma}_i$ is the mean dispersion within the spectral region around pixel $i$. 

In case of several continuum regions (when fitting multiple lines), this formalism can be applied to the set of $(\alpha_1, \ldots, \alpha_n)$ describing piece-wise correction to the continuum. In that case, we will have a product of $p(\alpha_k|d_k)$ in Eqs.~\eqref{eq:lnL_modified} and \eqref{eq:lnL_marginalize} (with marginalisation over each $\alpha_k$). To minimise the number of parameters it is reasonable to assume that all $p(\alpha_k|d_k)$ have the same shape with $d_1=\ldots=d_n\equiv d$, i.e. we are left with only one parameter $d$ that is used to describe the sampling of the continuum uncertainty. In that case we obtain a similar likelihood function
\begin{equation}
    p(D|\theta, d) = p(D|\theta) \cdot \prod\limits_{k=1}^{n}\frac1{\sqrt{2\pi A_k}d} \exp\left(\frac{B_k^2}{2A_k}\right),
\end{equation}
where $A_k$ and $B_k$ expressed by the similar expressions as in Eq.~\ref{eq:A} and Eq.~\ref{eq:B} with the exception that the sums are taken over the pixels corresponding to $k^{\rm th}$ interval of the continuum correction. 

Following this approach, we fitted the \ion{Fe}{i} lines at $z=0.45$ towards HE\,0001$-$2340. We assume flat priors on the $z$, $b$, $\log N$ and covering fraction $C_f$. For continuum placement uncertainties we used one hyper-parameter, $h$, which as presented above, describes the dispersion of the continuum displacements distribution (we assumed it to be a Gaussian) in units of median pixel uncertainties in $\sim1 $\AA-wide region around each of the considered line. 
We used a prior on $h$ to be Gaussian with 1.0 mean and dispersion 0.2. The posterior probability function of the parameters was sampled using an affine-invariant Monte-Carlo Markov Chain sampler.

%% file: main.bbl
\begin{thebibliography}{69}
\expandafter\ifx\csname natexlab\endcsname\relax\def\natexlab#1{#1}\fi

\bibitem[{{Agafonova} {et~al.}(2011){Agafonova}, {Molaro}, {Levshakov}, \&
  {Hou}}]{Agafonova2011}
{Agafonova}, I.~I., {Molaro}, P., {Levshakov}, S.~A., \& {Hou}, J.~L. 2011,
  \aap, 529, A28

\bibitem[{{Balashev} {et~al.}(2019){Balashev}, {Klimenko}, {Noterdaeme},
  {Krogager}, {Varshalovich}, {Ivanchik}, {Petitjean}, {Srianand}, \&
  {Ledoux}}]{Balashev2019}
{Balashev}, S.~A., {Klimenko}, V.~V., {Noterdaeme}, P., {et~al.} 2019, \mnras,
  490, 2668

\bibitem[{{Balashev} \& {Noterdaeme}(2018)}]{Balashev2018}
{Balashev}, S.~A. \& {Noterdaeme}, P. 2018, \mnras, 478, L7

\bibitem[{{Balashev} {et~al.}(2017){Balashev}, {Noterdaeme}, {Rahmani},
  {Klimenko}, {Ledoux}, {Petitjean}, {Srianand}, {Ivanchik}, \&
  {Varshalovich}}]{Balashev2017}
{Balashev}, S.~A., {Noterdaeme}, P., {Rahmani}, H., {et~al.} 2017, \mnras, 470,
  2890

\bibitem[{{Balashev} {et~al.}(2011){Balashev}, {Petitjean}, {Ivanchik},
  {Ledoux}, {Srianand}, {Noterdaeme}, \& {Varshalovich}}]{Balashev2011}
{Balashev}, S.~A., {Petitjean}, P., {Ivanchik}, A.~V., {et~al.} 2011, \mnras,
  418, 357

\bibitem[{{Balashev} {et~al.}(2016){Balashev}, {Zavarygin}, {Ivanchik},
  {Telikova}, \& {Varshalovich}}]{Balashev2016}
{Balashev}, S.~A., {Zavarygin}, E.~O., {Ivanchik}, A.~V., {Telikova}, K.~N., \&
  {Varshalovich}, D.~A. 2016, \mnras, 458, 2188

\bibitem[{{Bergeron} \& {Boiss{\'e}}(2017)}]{Bergeron2017}
{Bergeron}, J. \& {Boiss{\'e}}, P. 2017, \aap, 604, A37

\bibitem[{{Boiss{\'e}} {et~al.}(2015){Boiss{\'e}}, {Bergeron}, {Prochaska},
  {P{\'e}roux}, \& {York}}]{Boisse2015}
{Boiss{\'e}}, P., {Bergeron}, J., {Prochaska}, J.~X., {P{\'e}roux}, C., \&
  {York}, D.~G. 2015, \aap, 581, A109

\bibitem[{{Bouret} {et~al.}(2018){Bouret}, {Neiner}, {G{\'o}mez de Castro},
  {Evans}, {Gaensicke}, {Shore}, {Fossati}, {Gry}, {Charlot}, {Marin},
  {Noterdaeme}, \& {Chaufray}}]{Bouret2018}
{Bouret}, J.-C., {Neiner}, C., {G{\'o}mez de Castro}, A.~I., {et~al.} 2018, in
  Society of Photo-Optical Instrumentation Engineers (SPIE) Conference Series,
  Vol. 10699, Space Telescopes and Instrumentation 2018: Ultraviolet to Gamma
  Ray, ed. J.-W.~A. {den Herder}, S.~{Nikzad}, \& K.~{Nakazawa}, 106993B

\bibitem[{{Burles} \& {Tytler}(1998)}]{Burles1998}
{Burles}, S. \& {Tytler}, D. 1998, \apj, 507, 732

\bibitem[{{Carswell} {et~al.}(2012){Carswell}, {Becker}, {Jorgenson}, {Murphy},
  \& {Wolfe}}]{Carswell2012}
{Carswell}, R.~F., {Becker}, G.~D., {Jorgenson}, R.~A., {Murphy}, M.~T., \&
  {Wolfe}, A.~M. 2012, \mnras, 422, 1700

\bibitem[{{Carswell} \& {Webb}(2014)}]{Carswell2014}
{Carswell}, R.~F. \& {Webb}, J.~K. 2014, {VPFIT: Voigt profile fitting program}

\bibitem[{{Cooke} {et~al.}(2014){Cooke}, {Pettini}, {Jorgenson}, {Murphy}, \&
  {Steidel}}]{Cooke2014}
{Cooke}, R.~J., {Pettini}, M., {Jorgenson}, R.~A., {Murphy}, M.~T., \&
  {Steidel}, C.~C. 2014, \apj, 781, 31

\bibitem[{{Cooke} {et~al.}(2018){Cooke}, {Pettini}, \& {Steidel}}]{Cooke2018}
{Cooke}, R.~J., {Pettini}, M., \& {Steidel}, C.~C. 2018, \apj, 855, 102

\bibitem[{{De Cia} {et~al.}(2018){De Cia}, {Ledoux}, {Petitjean}, \&
  {Savaglio}}]{DeCia2018}
{De Cia}, A., {Ledoux}, C., {Petitjean}, P., \& {Savaglio}, S. 2018, \aap, 611,
  A76

\bibitem[{{Dekker} {et~al.}(2000){Dekker}, {D'Odorico}, {Kaufer}, {Delabre}, \&
  {Kotzlowski}}]{Dekker2000}
{Dekker}, H., {D'Odorico}, S., {Kaufer}, A., {Delabre}, B., \& {Kotzlowski}, H.
  2000, in Proc. SPIE Vol. 4008, p. 534-545, Optical and IR Telescope
  Instrumentation and Detectors, Masanori Iye; Alan F. Moorwood; Eds., 534--545

\bibitem[{{D'Odorico}(2007)}]{Dodorico2007}
{D'Odorico}, V. 2007, \aap, 470, 523

\bibitem[{{Dumont} \& {Webb}(2017)}]{Dumont2017}
{Dumont}, V. \& {Webb}, J.~K. 2017, \mnras, 468, 1568

\bibitem[{{Dutta} {et~al.}(2014){Dutta}, {Srianand}, {Rahmani}, {Petitjean},
  {Noterdaeme}, \& {Ledoux}}]{Dutta2014}
{Dutta}, R., {Srianand}, R., {Rahmani}, H., {et~al.} 2014, \mnras, 440, 307

\bibitem[{{Field} {et~al.}(1969){Field}, {Goldsmith}, \& {Habing}}]{Field1969}
{Field}, G.~B., {Goldsmith}, D.~W., \& {Habing}, H.~J. 1969, \apjl, 155, L149

\bibitem[{{Froning} {et~al.}(2006){Froning}, {Osterman}, {Beasley}, {Green}, \&
  {Beland}}]{Hros2006}
{Froning}, C., {Osterman}, S., {Beasley}, M., {Green}, J., \& {Beland}, S.
  2006, in Society of Photo-Optical Instrumentation Engineers (SPIE) Conference
  Series, Vol. 6269, Society of Photo-Optical Instrumentation Engineers (SPIE)
  Conference Series, ed. I.~S. {McLean} \& M.~{Iye}, 62691V

\bibitem[{{Ganguly} {et~al.}(1999){Ganguly}, {Eracleous}, {Charlton}, \&
  {Churchill}}]{Ganguly1999}
{Ganguly}, R., {Eracleous}, M., {Charlton}, J.~C., \& {Churchill}, C.~W. 1999,
  \aj, 117, 2594

\bibitem[{{Heiles} \& {Troland}(2003)}]{Heiles2003}
{Heiles}, C. \& {Troland}, T.~H. 2003, \apj, 586, 1067

\bibitem[{{Jones} {et~al.}(2010){Jones}, {Misawa}, {Charlton}, {Mshar}, \&
  {Ferland}}]{Jones2010}
{Jones}, T.~M., {Misawa}, T., {Charlton}, J.~C., {Mshar}, A.~C., \& {Ferland},
  G.~J. 2010, \apj, 715, 1497

\bibitem[{{Jorgenson} {et~al.}(2014){Jorgenson}, {Murphy}, {Thompson}, \&
  {Carswell}}]{Jorgenson2014}
{Jorgenson}, R.~A., {Murphy}, M.~T., {Thompson}, R., \& {Carswell}, R.~F. 2014,
  \mnras, 443, 2783

\bibitem[{{Kanekar} {et~al.}(2014){Kanekar}, {Prochaska}, {Smette}, {Ellison},
  {Ryan-Weber}, {Momjian}, {Briggs}, {Lane}, {Chengalur}, {Delafosse}, {Grave},
  {Jacobsen}, \& {de Bruyn}}]{Kanekar2014}
{Kanekar}, N., {Prochaska}, J.~X., {Smette}, A., {et~al.} 2014, \mnras, 438,
  2131

\bibitem[{{Kausch} {et~al.}(2015){Kausch}, {Noll}, {Smette}, {Kimeswenger},
  {Barden}, {Szyszka}, {Jones}, {Sana}, {Horst}, \& {Kerber}}]{Kausch2015}
{Kausch}, W., {Noll}, S., {Smette}, A., {et~al.} 2015, \aap, 576, A78

\bibitem[{{Khaire} \& {Srianand}(2019)}]{Khaire2019}
{Khaire}, V. \& {Srianand}, R. 2019, \mnras, 484, 4174

\bibitem[{{Krogager} {et~al.}(2017){Krogager}, {M{\o}ller}, {Fynbo}, \&
  {Noterdaeme}}]{Krogager2017}
{Krogager}, J.~K., {M{\o}ller}, P., {Fynbo}, J.~P.~U., \& {Noterdaeme}, P.
  2017, \mnras, 469, 2959

\bibitem[{{Krogager} \& {Noterdaeme}(2020)}]{Krogager2020b}
{Krogager}, J.-K. \& {Noterdaeme}, P. 2020, \aap, 644, L6

\bibitem[{{Lampton} {et~al.}(1976){Lampton}, {Margon}, \&
  {Bowyer}}]{Lampton1976}
{Lampton}, M., {Margon}, B., \& {Bowyer}, S. 1976, \apj, 208, 177

\bibitem[{{Ledoux} {et~al.}(2006){Ledoux}, {Petitjean}, {Fynbo}, {M{\o}ller},
  \& {Srianand}}]{Ledoux2006}
{Ledoux}, C., {Petitjean}, P., {Fynbo}, J.~P.~U., {M{\o}ller}, P., \&
  {Srianand}, R. 2006, \aap, 457, 71

\bibitem[{{Ledoux} {et~al.}(2003){Ledoux}, {Petitjean}, \&
  {Srianand}}]{Ledoux2003}
{Ledoux}, C., {Petitjean}, P., \& {Srianand}, R. 2003, \mnras, 346, 209

\bibitem[{{Ledoux} {et~al.}(2002){Ledoux}, {Srianand}, \&
  {Petitjean}}]{Ledoux2002}
{Ledoux}, C., {Srianand}, R., \& {Petitjean}, P. 2002, \aap, 392, 781

\bibitem[{{Lee} {et~al.}(2020){Lee}, {Webb}, {Carswell}, \&
  {Milakovic}}]{Lee2020}
{Lee}, C.-C., {Webb}, J.~K., {Carswell}, R.~F., \& {Milakovic}, D. 2020, arXiv
  e-prints, arXiv:2008.02583

\bibitem[{{Levshakov} {et~al.}(2009){Levshakov}, {Agafonova}, {Molaro},
  {Reimers}, \& {Hou}}]{Levshakov2009}
{Levshakov}, S.~A., {Agafonova}, I.~I., {Molaro}, P., {Reimers}, D., \& {Hou},
  J.~L. 2009, \aap, 507, 209

\bibitem[{{Marconi} {et~al.}(2021){Marconi}, {Abreu}, {Adibekyan}, {Aliverti},
  {Allende Prieto}, {Amado}, {Amate}, {Artigau}, {Augusto}, {Barros},
  {Becerril}, {Benneke}, {Bergin}, {Berio}, {Bezawada}, {Boisse}, {Bonfils},
  {Bouchy}, {Broeg}, {Cabral}, {Calvo-Ortega}, {Canto Martins}, {Chazelas},
  {Chiavassa}, {Christensen}, {Cirami}, {Coretti}, {Covino}, {Cresci},
  {Cristiani}, {Cunha Parro}, {Cupani}, {de Castro Le{\~a}o}, {Renan de
  Medeiros}, {Furlande Souza}, {Di Marcantonio}, {Di Varano}, {D'Odorico},
  {Doyon}, {Drass}, {Figueira}, {Belen Fragoso}, {Uldall Fynbo}, {Gallo},
  {Genoni}, {Gonz{\'a}lez Hern{\'a}ndez}, {Haehnelt}, {Hlavacek-Larrondo},
  {Hughes}, {Huke}, {Humphrey}, {Kjeldsen}, {Korn}, {Kouach}, {Landoni},
  {Liske}, {Lovis}, {Lunney}, {Maiolino}, {Malo}, {Marquart}, {Martins},
  {Mason}, {Molaro}, {Monnier}, {Monteiro}, {Mordasini}, {Morris},
  {Mucciarelli}, {Murray}, {Niedzielski}, {Nunes}, {Oliva}, {Origlia},
  {Pall{\'e}}, {Pariani}, {Parr-Burman}, {Pe{\~n}ate}, {Pepe}, {Pinna},
  {Piskunov}, {Rasilla Pi{\~n}eiro}, {Rebolo}, {Rees}, {Reiners}, {Riva},
  {Romano}, {Rousseau}, {Sanna}, {Santos}, {Sarajlic}, {Shen}, {Sortino},
  {Sosnowska}, {Sousa}, {Stempels}, {Strassmeier}, {Tenegi}, {Tozzi}, {Udry},
  {Valenziano}, {Vanzi}, {Weber}, {Woche}, {Xompero}, {Zackrisson}, \&
  {Zapatero Osorio}}]{Marconi2021}
{Marconi}, A., {Abreu}, M., {Adibekyan}, V., {et~al.} 2021, The Messenger, 182,
  27

\bibitem[{{Milakovi{\'c}} {et~al.}(2021){Milakovi{\'c}}, {Lee}, {Carswell},
  {Webb}, {Molaro}, \& {Pasquini}}]{Milakovic2021}
{Milakovi{\'c}}, D., {Lee}, C.-C., {Carswell}, R.~F., {et~al.} 2021, \mnras,
  500, 1

\bibitem[{{Murphy} \& {Berengut}(2014)}]{Murphy2014}
{Murphy}, M.~T. \& {Berengut}, J.~C. 2014, \mnras, 438, 388

\bibitem[{{Murphy} {et~al.}(2008){Murphy}, {Webb}, \& {Flambaum}}]{Murphy2008}
{Murphy}, M.~T., {Webb}, J.~K., \& {Flambaum}, V.~V. 2008, \mnras, 384, 1053

\bibitem[{{Neeleman} {et~al.}(2015){Neeleman}, {Prochaska}, \&
  {Wolfe}}]{Neeleman2015}
{Neeleman}, M., {Prochaska}, J.~X., \& {Wolfe}, A.~M. 2015, \apj, 800, 7

\bibitem[{{Noterdaeme} {et~al.}(2017){Noterdaeme}, {Krogager}, {Balashev},
  {Ge}, {Gupta}, {Kr{\"u}hler}, {Ledoux}, {Murphy}, {P{\^a}ris}, {Petitjean},
  {Rahmani}, {Srianand}, \& {Ubachs}}]{Noterdaeme2017}
{Noterdaeme}, P., {Krogager}, J.-K., {Balashev}, S., {et~al.} 2017, \aap, 597,
  A82

\bibitem[{{Noterdaeme} {et~al.}(2008){Noterdaeme}, {Ledoux}, {Petitjean}, \&
  {Srianand}}]{Noterdaeme2008b}
{Noterdaeme}, P., {Ledoux}, C., {Petitjean}, P., \& {Srianand}, R. 2008, \aap,
  481, 327

\bibitem[{{Noterdaeme} {et~al.}(2012){Noterdaeme}, {L{\'o}pez}, {Dumont},
  {Ledoux}, {Molaro}, \& {Petitjean}}]{Noterdaeme2012a}
{Noterdaeme}, P., {L{\'o}pez}, S., {Dumont}, V., {et~al.} 2012, \aap, 542, L33

\bibitem[{{Noterdaeme} {et~al.}(2007){Noterdaeme}, {Petitjean}, {Srianand},
  {Ledoux}, \& {Le Petit}}]{Noterdaeme2007}
{Noterdaeme}, P., {Petitjean}, P., {Srianand}, R., {Ledoux}, C., \& {Le Petit},
  F. 2007, \aap, 469, 425

\bibitem[{{Noterdaeme} {et~al.}(2011){Noterdaeme}, {Petitjean}, {Srianand},
  {Ledoux}, \& {L{\'o}pez}}]{Noterdaeme2011}
{Noterdaeme}, P., {Petitjean}, P., {Srianand}, R., {Ledoux}, C., \&
  {L{\'o}pez}, S. 2011, \aap, 526, L7+

\bibitem[{{O'Meara} {et~al.}(2006){O'Meara}, {Burles}, {Prochaska}, {Prochter},
  {Bernstein}, \& {Burgess}}]{OMeara2006}
{O'Meara}, J.~M., {Burles}, S., {Prochaska}, J.~X., {et~al.} 2006, \apjl, 649,
  L61

\bibitem[{{O'Meara} {et~al.}(2001){O'Meara}, {Tytler}, {Kirkman}, {Suzuki},
  {Prochaska}, {Lubin}, \& {Wolfe}}]{O'Meara2001}
{O'Meara}, J.~M., {Tytler}, D., {Kirkman}, D., {et~al.} 2001, \apj, 552, 718

\bibitem[{{Pepe} {et~al.}(2021){Pepe}, {Cristiani}, {Rebolo}, {Santos},
  {Dekker}, {Cabral}, {Di Marcantonio}, {Figueira}, {Lo Curto}, {Lovis},
  {Mayor}, {M{\'e}gevand}, {Molaro}, {Riva}, {Zapatero Osorio}, {Amate},
  {Manescau}, {Pasquini}, {Zerbi}, {Adibekyan}, {Abreu}, {Affolter}, {Alibert},
  {Aliverti}, {Allart}, {Allende Prieto}, {{\'A}lvarez}, {Alves}, {Avila},
  {Baldini}, {Bandy}, {Barros}, {Benz}, {Bianco}, {Borsa}, {Bourrier},
  {Bouchy}, {Broeg}, {Calderone}, {Cirami}, {Coelho}, {Conconi}, {Coretti},
  {Cumani}, {Cupani}, {D'Odorico}, {Damasso}, {Deiries}, {Delabre},
  {Demangeon}, {Dumusque}, {Ehrenreich}, {Faria}, {Fragoso}, {Genolet},
  {Genoni}, {G{\'e}nova Santos}, {Gonz{\'a}lez Hern{\'a}ndez}, {Hughes},
  {Iwert}, {Kerber}, {Knudstrup}, {Landoni}, {Lavie}, {Lillo-Box}, {Lizon},
  {Maire}, {Martins}, {Mehner}, {Micela}, {Modigliani}, {Monteiro}, {Monteiro},
  {Moschetti}, {Murphy}, {Nunes}, {Oggioni}, {Oliveira}, {Oshagh}, {Pall{\'e}},
  {Pariani}, {Poretti}, {Rasilla}, {Rebord{\~a}o}, {Redaelli}, {Santana
  Tschudi}, {Santin}, {Santos}, {S{\'e}gransan}, {Schmidt}, {Segovia},
  {Sosnowska}, {Sozzetti}, {Sousa}, {Span{\`o}}, {Su{\'a}rez Mascare{\~n}o},
  {Tabernero}, {Tenegi}, {Udry}, \& {Zanutta}}]{Pepe2021}
{Pepe}, F., {Cristiani}, S., {Rebolo}, R., {et~al.} 2021, \aap, 645, A96

\bibitem[{{P{\'e}roux} \& {Howk}(2020)}]{Peroux2020}
{P{\'e}roux}, C. \& {Howk}, J.~C. 2020, \araa, 58, 363

\bibitem[{{Petitjean} {et~al.}(2008){Petitjean}, {Ledoux}, \&
  {Srianand}}]{Petitjean2008}
{Petitjean}, P., {Ledoux}, C., \& {Srianand}, R. 2008, \aap, 480, 349

\bibitem[{{Pettini} \& {Cooke}(2012)}]{Pettini2012}
{Pettini}, M. \& {Cooke}, R. 2012, \mnras, 425, 2477

\bibitem[{{Pettini} {et~al.}(2008){Pettini}, {Zych}, {Murphy}, {Lewis}, \&
  {Steidel}}]{Pettini2008}
{Pettini}, M., {Zych}, B.~J., {Murphy}, M.~T., {Lewis}, A., \& {Steidel}, C.~C.
  2008, \mnras, 391, 1499

\bibitem[{{Prochaska} \& {Wolfe}(1997)}]{Prochaska1997}
{Prochaska}, J.~X. \& {Wolfe}, A.~M. 1997, \apj, 487, 73

\bibitem[{{Prochaska} {et~al.}(2007){Prochaska}, {Wolfe}, {Howk}, {Gawiser},
  {Burles}, \& {Cooke}}]{Prochaska2007}
{Prochaska}, J.~X., {Wolfe}, A.~M., {Howk}, J.~C., {et~al.} 2007, \apjs, 171,
  29

\bibitem[{{Rahmani} {et~al.}(2013){Rahmani}, {Wendt}, {Srianand}, {Noterdaeme},
  {Petitjean}, {Molaro}, {Whitmore}, {Murphy}, {Centurion}, {Fathivavsari},
  {D'Odorico}, {Evans}, {Levshakov}, {Lopez}, {Martins}, {Reimers}, \&
  {Vladilo}}]{Rahmani2013}
{Rahmani}, H., {Wendt}, M., {Srianand}, R., {et~al.} 2013, \mnras, 435, 861

\bibitem[{{Richter} {et~al.}(2005){Richter}, {Ledoux}, {Petitjean}, \&
  {Bergeron}}]{Richter2005}
{Richter}, P., {Ledoux}, C., {Petitjean}, P., \& {Bergeron}, J. 2005, \aap,
  440, 819

\bibitem[{{Salpeter}(1976)}]{Salpeter1976}
{Salpeter}, E.~E. 1976, \apj, 206, 673

\bibitem[{{Smette} {et~al.}(2015){Smette}, {Sana}, {Noll}, {Horst}, {Kausch},
  {Kimeswenger}, {Barden}, {Szyszka}, {Jones}, {Gallenne}, {Vinther},
  {Ballester}, \& {Taylor}}]{Smette2015}
{Smette}, A., {Sana}, H., {Noll}, S., {et~al.} 2015, \aap, 576, A77

\bibitem[{{Srianand} {et~al.}(2012){Srianand}, {Gupta}, {Petitjean},
  {Noterdaeme}, {Ledoux}, {Salter}, \& {Saikia}}]{Srianand2012}
{Srianand}, R., {Gupta}, N., {Petitjean}, P., {et~al.} 2012, \mnras, 421, 651

\bibitem[{{Srianand} {et~al.}(2005){Srianand}, {Petitjean}, {Ledoux},
  {Ferland}, \& {Shaw}}]{Srianand2005}
{Srianand}, R., {Petitjean}, P., {Ledoux}, C., {Ferland}, G., \& {Shaw}, G.
  2005, \mnras, 362, 549

\bibitem[{{Szentgyorgyi} {et~al.}(2016){Szentgyorgyi}, {Baldwin}, {Barnes},
  {Bean}, {Ben-Ami}, {Brennan}, {Budynkiewicz}, {Chun}, {Conroy}, {Crane},
  {Epps}, {Evans}, {Evans}, {Foster}, {Frebel}, {Gauron}, {Guzm{\'a}n}, {Hare},
  {Jang}, {Jang}, {Jordan}, {Kim}, {Kim}, {Mendes de Oliveira},
  {Lopez-Morales}, {McCracken}, {McMuldroch}, {Miller}, {Mueller}, {Oh},
  {Onyuksel}, {Ordway}, {Park}, {Park}, {Park}, {Paxson}, {Phillips},
  {Plummer}, {Podgorski}, {Seifahrt}, {Stark}, {Steiner}, {Uomoto},
  {Walsworth}, \& {Yu}}]{Gclef2016}
{Szentgyorgyi}, A., {Baldwin}, D., {Barnes}, S., {et~al.} 2016, in Society of
  Photo-Optical Instrumentation Engineers (SPIE) Conference Series, Vol. 9908,
  Ground-based and Airborne Instrumentation for Astronomy VI, ed. C.~J.
  {Evans}, L.~{Simard}, \& H.~{Takami}, 990822

\bibitem[{{Vangioni} \& {Olive}(2019)}]{Vangioni2019}
{Vangioni}, E. \& {Olive}, K.~A. 2019, \mnras, 484, 3561

\bibitem[{{Vladilo} {et~al.}(2018){Vladilo}, {Gioannini}, {Matteucci}, \&
  {Palla}}]{Vladilo2018}
{Vladilo}, G., {Gioannini}, L., {Matteucci}, F., \& {Palla}, M. 2018, \apj,
  868, 127

\bibitem[{{Vogt} {et~al.}(1994){Vogt}, {Allen}, {Bigelow}, {Bresee}, {Brown},
  {Cantrall}, {Conrad}, {Couture}, {Delaney}, {Epps}, {Hilyard}, {Hilyard},
  {Horn}, {Jern}, {Kanto}, {Keane}, {Kibrick}, {Lewis}, {Osborne},
  {Pardeilhan}, {Pfister}, {Ricketts}, {Robinson}, {Stover}, {Tucker}, {Ward},
  \& {Wei}}]{Vogt1994}
{Vogt}, S.~S., {Allen}, S.~L., {Bigelow}, B.~C., {et~al.} 1994, in Society of
  Photo-Optical Instrumentation Engineers (SPIE) Conference Series, Vol. 2198,
  Instrumentation in Astronomy VIII, ed. D.~L. {Crawford} \& E.~R. {Craine},
  362

\bibitem[{{Vreeswijk} {et~al.}(2007){Vreeswijk}, {Ledoux}, {Smette}, {Ellison},
  {Jaunsen}, {Andersen}, {Fruchter}, {Fynbo}, {Hjorth}, {Kaufer}, {M{\o}ller},
  {Petitjean}, {Savaglio}, \& {Wijers}}]{Vreeswijk2007}
{Vreeswijk}, P.~M., {Ledoux}, C., {Smette}, A., {et~al.} 2007, \aap, 468, 83

\bibitem[{{Welsh} {et~al.}(2020){Welsh}, {Cooke}, {Fumagalli}, \&
  {Pettini}}]{Welsh2020}
{Welsh}, L., {Cooke}, R., {Fumagalli}, M., \& {Pettini}, M. 2020, \mnras, 494,
  1411

\bibitem[{{Whitmore} \& {Murphy}(2015)}]{Whitmore2015}
{Whitmore}, J.~B. \& {Murphy}, M.~T. 2015, \mnras, 447, 446

\bibitem[{{Wolfire} {et~al.}(2003){Wolfire}, {McKee}, {Hollenbach}, \&
  {Tielens}}]{Wolfire2003}
{Wolfire}, M.~G., {McKee}, C.~F., {Hollenbach}, D., \& {Tielens}, A.~G.~G.~M.
  2003, \apj, 587, 278

\end{thebibliography}
